  \documentclass{aa}  

 \usepackage{natbib}
  \bibpunct{(}{)}{;}{a}{}{,}             %% natbib format for A&A and ApJ
 \usepackage{graphicx}
\usepackage{txfonts}

 \usepackage[breaklinks=true]{hyperref} %% to avoid \citeads line fills
 \hypersetup{colorlinks=true,urlcolor=blue,citecolor=blue,pdfborder= 0 0 0}

\begin{document}

   \title{Cosmic metal invaders: Intergalactic \ion{O}{VII} as a tracer of the warm-hot intergalactic medium within cosmic filaments in the EAGLE simulation}
    \titlerunning{Cosmic metal invaders}
%   \subtitle{I. Overviewing the $\kappa$-mechanism}

   \author{T. Tuominen
          \inst{1}\thanks{\email{tuominen@ut.ee}}
          \and
          J. Nevalainen\inst{1}%\fnmsep%\thanks{Just to show the usage
          %of the elements in the author field}
        \and
        P. Heinämäki\inst{2}
        \and 
        E. Tempel\inst{1,3}
            \and 
        N. Wijers\inst{4,5}
        \and    
        M. Bonamente \inst{6}
        \and
        M.A. Aragon-Calvo \inst{7}
        \and
        A. Finoguenov\inst{8}
          }

   \institute{Tartu Observatory, University of Tartu,
              61602 Tõravere, Tartumaa, Estonia
                      \and
             Tuorla Observatory, Department of Physics and Astronomy, University of Turku, 
             20014 Turku, Finland
             %\email{c.ptolemy@hipparch.uheaven.space}
             %\thanks{The university of heaven temporarily does not
             %        accept e-mails}
             \and
             Estonian Academy of Sciences, Kohtu 6, 10130 Tallinn, Estonia
             \and
             Leiden Observatory, Leiden University, PO Box 9513, NL-2300 RA Leiden, The Netherlands
             \and
             CIERA and Department of Physics and Astronomy, Northwestern University, 1800 Sherman Ave, Evanston, IL 60201, USA
             \and
              The University of Alabama in Huntsville, Huntsville, AL 35899, USA 
              \and
              Instituto de Astronom\'{i}a, UNAM, Apdo. Postal 106, Ensenada 22800, B.C., M\'{e}xico
              \and
              Department of Physics, University of Helsinki, Gustaf Hällströmin katu 2a,  Helsinki, FI-00014, Finland
             }

   \date{Received ; accepted }

% \abstract{}{}{}{}{} 
% 5 {} token are mandatory
 
  \abstract
   {%\LEt{ General notes: A.) I have edited to UK English spelling and grammar conventions. B.) A\&A uses the past tense to describe specific methods used in a paper and the present tense to describe general methods as well as findings, including the findings of recent papers (within the past ten or so years). Kindly make any necessary changes (I have made some, but my edits are by no means exhaustive in this respect). See Sect. 6 of the language guide https://www.aanda.org/for-authors/language-editing/6-verb-tenses.}
   The current observational status of the hot ($\log T(K) > 5.5$) intergalactic medium (IGM) remains incomplete. 
   While recent X-ray emission and Sunyaev-Zeldovich effect observations from stacking large numbers of Cosmic Web filaments 
   %as well as fast radio bursts
   have yielded 
   statistically significant detections of this phase, direct statistically significant measurements of single objects remain scarce. The lack of such a sample currently prevents a robust analysis of the cosmic baryon content composed of the hot IGM, which would potentially help solve the cosmological missing baryons problem.}
   {In order to improve the observationally challenging search for the missing baryons, we utilise the theoretical avenue afforded by the EAGLE simulations. Our aim is to get insights into the metal enrichment of the Cosmic Web and the distribution of highly ionised metals in the IGM. Our goal is to aid in the planning of future X-ray observations of the hot intergalactic plasma.}
   {We detected the filamentary %galaxy 
   network by applying the Bisous formalism to galaxies in the EAGLE simulation. We characterised the spatial distributions of oxygen and $\ion{O}{VII}$ and studied their mass and volume filling fractions in the filaments.  Since oxygen is formed in and expelled from galaxies, we also studied the surroundings of haloes. We used this information to construct maps of the $\ion{O}{VII}$ column density and determine the feasibility of detecting it via absorption  with Athena %\LEt{ In this paper you write both "Athena" and "ATHENA." Assuming they are the same, please choose one convention and use it consistently.\ Furthermore, if "ATHENA" is correct, consider defining the acronym in the main text.}
   X-IFU.}
   {Within EAGLE, the oxygen and $\ion{O}{VII}$ number densities drop dramatically beyond the virial radii of haloes. 
   %Thus, the amount and spread of the intergalactic metals remain low within Bisous filaments. 
   In the most favourable scenario, the median extent of \ion{O}{VII} above the Athena X-IFU detection limit is $\approx 700$ kpc. Since galaxies are relatively far apart from one another, %...and consequently 
   only $\sim 1\%$ of the filament volumes are filled with \ion{O}{VII} at high enough column densities to be detectable by X-IFU. 
    %This is contrary to the assumption that the \ion{O}{VII} is %well mixed with the intergalactic baryons and 
    %homogeneously distributed throughout the filaments, required for estimating the cosmic contribution of the 
     The highly non-homogeneous distribution of the detectable \ion{O}{VII} complicates the usage of the measurements of the intergalactic \ion{O}{VII} absorbers for tracing the missing baryons and estimating their contribution to the cosmic baryon budget. 
           Instead, the detectable volumes form narrow and dense %... kuinka paljon näissä OVII massaa..
      envelopes around haloes, while the rest of the \ion{O}{VII} is diluted at low 
      %median 
      densities within the full filament volumes. This localised nature, in turn, results in a low chance ($\sim$10-20\% per sight line)  of detecting intergalactic \ion{O}{VII} with Athena X-IFU within the observational SDSS catalogue of nearby filaments.
      % to anna interception probabilityn numeroarvoja... . 
   %However, already with a depth of 100 Mpc filaments cover over 90\% of the projected area. 
   Fortunately, with deeper filament samples, such as those provided via the future 4MOST 4HS survey, the chances of intercepting an absorbing system are expected to increase up to a comfortable level of $\sim 50\%$ per sight line.}  
   {Based on EAGLE results, 
   %even 
   targeting the Cosmic Web with Athena %indication is that targeting filaments may
may only result in tip-of-the-iceberg detections of the intergalactic \ion{O}{VII}, which is located in the galaxy outskirts. %\LEt{ Verify that your intended meaning has not been changed.} 
This would not be enough to conclusively solve the missing baryon problem. However, the projection of many filaments into a single line of sight will enable a useful X-ray observation strategy with Athena X-IFU for the hot cosmic baryon gas, reducing the amount of baryons still missing by up to $\sim 25\%$. %{\bf How much? PRIORITY 1}. %SIMBA...

   }


   \maketitle

\section{Introduction}
\label{intro}
The concordance cosmology model and \textit{Planck} measurements provide very tight constraints for the cosmic baryon density \citep{2014A&A...571A...1P}.
When estimating the contributions of different observed baryon components,  about half of the expected cosmic baryon budget has until recently remained unaccounted for \citep{Bregman07, 2012ApJ...759...23S}.
This discrepancy has been dubbed the missing baryons problem. In recent years, progress has been made towards the detection of these missing baryons through a wide range of methods. Stacking the \textit{Planck} \citep{2016A&A...594A..22P} thermal Sunyaev-Zeldovich (tSZ) maps between
a large number of cluster pairs, \citet{2019A&A...624A..48D} and \citet{2019MNRAS.483..223T} discovered an excess signal attributed to filaments or `bridges' between the clusters. In a similar vein, \citet{2020A&A...637A..41T} detected an excess in the \textit{Planck} tSZ signal by stacking a large number of filaments identified with the DisPerSe filament finding method \citep{2011MNRAS.414..350S}, inferring the baryon density and temperature of said filaments. Stacking the same filaments with ROSAT \citep{1997ApJ...485..125S} background maps, \citet{2020A&A...643L...2T} discovered an X-ray emission signal corresponding to the hot intergalactic medium (IGM). Recently, the dispersion measures of localised fast radio bursts with known redshifts have provided quantitative observational estimates for intergalactic baryon densities \citep{2020Natur.581..391M}. These estimates fill the missing baryon gap and thus may provide one solution to the missing baryon problem. In this work we focus on another solution, the direct X-ray absorption detection of the hot phase of the intergalactic diffuse matter, which we define as missing baryons for the purpose of this paper.
%\footnote{We do not discuss in this work the problem of measurements yielding smaller than cosmic baryon to dark matter density within r200 in galaxies}. 
%The accounting of the full observational contribution to the cosmic baryon density is a complicated task.
 
Simulations \citep[e.g.][]{1999ApJ...514....1C,2001ApJ...552..473D,2006MNRAS.370..656D,2012MNRAS.425.1640T,2019MNRAS.486.3766M,2021A&A...649A.117G} suggest that the baryon component that had until recently avoided detection consists of the cosmic diffuse baryons in the Cosmic Web.
Since the problem is likely due to current observational limits, we divide the diffuse baryonic structure of the Universe into two observationally different domains: 
(1) the more easily detectable high-density galaxy-scale ($\sim$ 100 kpc) environment and (2) the observationally more challenging low-density large-scale ($\sim 1-10$ Mpc) intergalactic space in the Cosmic Web filaments. Consequently, we propose separating the missing baryon problem into two parts: %Part 1
(1) the galactic missing baryons, related to baryons bound to the galaxies, and %Part 2
(2) the cosmic missing baryons, related to the intergalactic volumes within the Cosmic Web filaments.%\LEt{ A\&A does not allow the use of italics for emphasis or to denote a special meaning. If you want to mark a special meaning of words or phrases, you can use quotation marks the first time a word/phrase appears.}

Regarding the missing galactic baryons, the Evolution and Assembly of GaLaxies and their Environments (EAGLE) simulations have been used to investigate the detection feasibility of the missing baryons within galaxies via X-ray absorption \citep{2020MNRAS.498..574W} and emission \citep{Wijers:22}. The results indicate that at sight lines passing through a halo of $\log $M/M$_{\odot} = 12.0-13.5$ at a radial distance comparable to the halo's virial radius, R$_{200}$,
the column density of the widely used \ion{O}{VII} ion is at the level of $\log N_{\ion{O}{VII}} ($cm$^{-2}) \sim 15$.
The predicted level is only achievable with the very rare megasecond %\LEt{ If this is a unit, please write it out.}
observations with the currently best instrument for this, the \textit{XMM-Newton}/RGS. %\LEt{ Consider defining.}.
Typical 100 ks observations reach only a fraction of this column density level and are thus limited to the central regions. Therefore, a large fraction of the expected baryons in the outer regions of galaxies remain unobserved. 
%This is Part 1 of the observational problem of the missing baryons, i.e. the one related to galaxies. 
According to the EAGLE simulations, future instruments such as the Athena X-ray Integral Field Unit (X-IFU) %\LEt{ Consider defining.}
have the potential to efficiently probe the circumgalactic medium (CGM) within the virial radii of galaxies via \ion{O}{VII} absorption and emission \citep{2020MNRAS.498..574W,Wijers:22}.  
This may enable us to solve to the galactic missing baryons problem.%Part 1

Cosmological simulations indicate that galaxies (within R$_{200}$) occupy only $\approx$ 1\%  of the volume covered by the Cosmic Web filaments \citep[e.g.][]{2021A&A...646A.156T}. However, due to the orders-of-magnitude higher density of the diffuse baryons in galaxies, compared to that in the filaments (overdensity $\sim 10$), their contribution to the cosmic baryon budget is significant. Namely, $\approx$ 25\% of the baryon mass contained within the full volumes of the cosmic filaments are concentrated within virial radii of haloes in the EAGLE simulations. Yet, most of the cosmic baryons are located outside R$_{200}$. Thus, even if the relatively dense galactic environment were one day to be robustly sampled observationally, the cosmic missing baryon problem would still remain unsolved. One cannot use the tips of icebergs to measure the mass of the ice of the sea directly, and similarly one cannot use the measurements of baryons within R$_{200}$ to derive the observational contribution of the intergalactic Cosmic Web volumes outside R$_{200}$ to the cosmic baryon budget.\footnote{One can make assumptions about the distribution of the hot gas in the CGM and in the intergalactic space in the filaments, but an assumption is not a solution to the observational missing baryons problem.}

In this work we focus on the cosmic missing baryon problem, in particular on the prospects of detecting the missing baryons in the cosmic filaments via X-ray absorption.
According to simulations, a significant fraction of the intergalactic cosmic baryons reside in the hot phase ($\log T(K) \ge 5.5$) of the warm-hot intergalactic medium (WHIM), corresponding to the X-ray regime. For example, our previous work on EAGLE \citep{2021A&A...646A.156T} yielded that $\approx 29\%$ of the EAGLE baryon budget is intergalactic hot WHIM, while the SIMBA %\LEt{ Consider defining.}
simulation \citep{Bradley:2022}, due to its stronger active galactic nucleus (AGN) jets, gives a bigger fraction of $\sim$70\% for the whole WHIM phase.

The spread of metals into the Cosmic Web relies on the galactic winds driven by supernovae (SNe) and AGN jets. Assuming that the IGM is efficiently enriched by metals, the densities of ions such as \ion{O}{VII} and \ion{O}{VIII} could be high enough to be detectable with X-rays. These ions could then be used to trace the hot WHIM. %galactic winds driven by supernovae (SNe) and AGN jets enrich the Cosmic Web with metals to high enough levels of mixing with WHIM, volume filling, and column density, the hot intergalactic WHIM may be traceable with high ions such as \ion{O}{VII} and \ion{O}{VIII} in X-rays.\LEt{  I don't understand this sentence.\ Please rephrase.}
In fact, another work based on the EAGLE simulations \citep{2022MNRAS.511.2600M} finds that the mass of the metals ejected from galaxies is substantial; it is comparable to that locked in stars at $z = 0$.  The work also finds that most of the ejected metal mass for haloes less massive than M $< 10^{13} $M$_{\odot}$ is located outside R$_{200}$ at $z = 0$. Tracing the SN-ejected particles in the EAGLE simulations, \citet{Kelly:22} find that $\sim$ 35\% of the baryons that during cosmic history have been located within the primary halo have %, in EAGLE, 
been ejected outside $R_{200}$ at $z = 0$. For Milky Way-like galaxies, the SN-ejected particles have been expelled to distances of 0.5-1.5 Mpc from the galactic centre. These results  suggest the existence of metal-enriched large-scale outflows.
We further investigate the ionisation of this component and the distribution of the consequent \ion{O}{VII} in the filaments of the Cosmic Web.

Assuming (1) a typical WHIM overdensity of the order of 10 in the filaments and (2) a WHIM temperature $\log T(K) \sim 6$, 
%(both consistent with the dense filament sample in the EAGLE simulations in Tuominen+2021), 
it follows that the commonly assumed oxygen abundance of 0.1 solar would yield column densities reaching the level of $\log N_{\ion{O}{VII}}($cm$^{-2}) = 15 $ for typical filament path lengths of the order of 1 Mpc. Since the path lengths through filaments are a factor of 10-100 larger than those through galaxies, the lower densities in filaments may be sufficient to bring the signal above the detection limit of current X-ray instruments in the most optimal situations.

The observational status of intergalactic \ion{O}{VII} is poor \citep[see][for recent reviews]{2019A&A...621A..88N, Nicastro22}.   
Currently, there is no large enough observational sample of intergalactic X-ray ion absorption line measurements of high significance for a meaningful characterisation of the hot WHIM absorber population. This is required to estimate the hot WHIM contribution to the cosmic baryon budget.
Because it contains a significant fraction of the cosmic baryons in simulations, and has a poor observational status, the hot WHIM fits the role of the cosmological missing baryons well. 

Recent works \citep{2019MNRAS.488.2947W,2020MNRAS.498..574W,Wijers:22} have presented an overview of the metal ion distribution in all environments within the EAGLE simulation, as well as a thorough study of the metal ions within the haloes. In this work we take a deeper look into the intergalactic metal ions in EAGLE%extend the EAGLE metal ion work into the intergalactic space
, in particular within the IGM in cosmic filaments. \citet{2019MNRAS.486.3766M} studied the metal ions in the filaments  in the Illustris-TNG %\LEt{ Consider defining.} 
simulations, finding 0.1 solar metallicity.
%and $\log f(N_{OVII}) = -14$ at $\log N_{OVII} (cm^{-2} = 15$ %(what does this mean?) 
We apply a very different approach in the current work.
We build on the work of \citet{2021A&A...646A.156T}, where we characterised the spatial distribution of the hot WHIM and its thermodynamic properties in the cosmic filaments in the EAGLE simulation. 
In that work we find that the WHIM in the central regions of the densest filaments reaches baryon overdensity levels of $\sim 10$ and temperatures $\log T(K) \sim 6$, as required for the X-ray detection (see above).    
In the present work we focus on oxygen, the most common metal in the Universe, and \ion{O}{VII}, the most abundant X-ray ion in typical filament environments. By investigating their spatial distributions within the cosmic filaments, we evaluate the column densities in order to derive observational predictions for near-future X-ray instrumentation.

Throughout this work we use $\Omega_{m}$ = 0.31, $\Omega_{\Lambda}$ = 0.69, $\Omega_{b}$ = 0.048, H$_{0}$ = 67.8 km s$^{-1}$ Mpc$^{-1}$, and $\sigma_{8}$ = 0.83, given by \citet{2014A&A...571A...1P}. These are the same values used in the EAGLE simulations. Since our interest lies in the nearby Universe at redshift $z$ = 0, all comoving and proper distances are equal. Thus, we present all distances simply in units of Mpc or kpc.

\section{Data}

\subsection{EAGLE simulation}
In this work we used the hydrodynamical EAGLE simulation \citep{2015MNRAS.446..521S,2015MNRAS.450.1937C,mcalpine_helly_etal_2016}. In particular, we used the largest simulation run, RefL0100N1504. It followed $1504^{3}$ gas and dark matter particles with initial masses of $m_{\mathrm{b}} = 1.81 \times 10^{6} $ M$_{\odot}$ and $m_{\mathrm{DM}} = 9.7 \times 10^{6} $ M$_{\odot}$, respectively, within a volume of $100^{3}$ Mpc$^{3}$. The $N-$body simulation code implemented in EAGLE is a modified version of Gadget3 \citep{2005MNRAS.364.1105S}, with adjustments to the smoothed particle hydrodynamics (SPH) as described in \citet{2015MNRAS.454.2277S}. 

In order to study the baryons within the local Universe, we selected the SPH particles within the snapshot corresponding to redshift $z=0$. In particular, we studied the smoothed oxygen abundances carried by each SPH particle \citep[see][for a description of the smoothed metallicities]{2009MNRAS.399..574W}. In addition, following \citet{2019MNRAS.488.2947W}, we combined the simulated oxygen fraction, temperature and baryon density together with tabulated ion fractions to obtain the number of \ion{O}{VII} ions within each SPH particle. The ion tables adopted here were created by \citet{2010MNRAS.407..544B} with the use of CLOUDY \citep{1998PASP..110..761F}. The tables include the effects of collisional ionisation, as well as photoionisation by a spatially uniform, redshift-dependent ultraviolet (UV) and X-ray background \citep{2001cghr.confE..64H}. This is consistent with the radiative cooling model used in EAGLE \citep{2009MNRAS.393...99W}.

\subsection{Intergalactic oxygen}
\label{ejected}
Since our focus was on the IGM, we needed to draw a border between the galaxies and intergalactic space. The problem is that this procedure is conceptually not well defined. Our practical approach was to utilise the friend-of-friends (FoF) procedure, already applied to the publicly available EAGLE data, which determines whether a given particle is bound to a dark matter halo or not. We considered the non-halo particles as intergalactic in this work. 

In more detail, we are interested in the intergalactic metals ejected from galaxies via galactic super-winds and AGN.
On the other hand, there are low-metal inflows affecting the metal budget in the outskirts of galaxies.
We do not attempt to separate these components. Rather, our pragmatic goal is to characterise the observationally relevant end result of those processes, that is, the total oxygen distribution outside the haloes at $z = 0$. %\LEt{ We do not allow the use of "e.g." or "i.e." within the main text (in parentheses or within figure/table captions is fine).}
This is sensitive to the implementation of the sub-grid physics, which varies between different simulations. We plan to investigate this issue in a future work by repeating the analyses presented in this work with other simulations with substantially different sub-grid implementations.

Our aim was to study the detection prospects of the intergalactic \ion{O}{VII}, done in terms of column densities. We also wanted to probe the three-dimensional information like volume filling fractions provided by EAGLE and this is done in terms of ion number densities. Thus, we needed an order of magnitude estimate for the number density level corresponding to the column density sensitivity limit of current and near-future X-ray instruments.
The \textit{XMM-Newton}/RGS is currently the most sensitive X-ray instrument for the \ion{O}{VII} absorption measurements. The $\sim$ 2\% uncertainty in the calibration of the effective area of the RGS \citep{2017AN....338..146K} sets the systematic floor for absorption line searches with this instrument. Assuming that a calibration problem causes a spurious 2\% dip in the continuum flux within one RGS resolution element (60 mÅ), the feature may be misinterpreted as an astrophysical line with an equivalent width of a few mÅ \citep{Nevalainen17}. This corresponds to an unresolved Gaussian line with $\log N$ (cm$^{-2}) \sim 15$. 
Thus, systematic uncertainties of the RGS effective area calibration do not permit the detection of \ion{O}{VII} column densities smaller than $\log N_\mathrm{\ion{O}{VII}}$ (cm$^{-2}) = 15$. On the observational side, the exceptionally long ($>$ Ms) exposures of the brightest blazars with RGS have reached the photon statistics corresponding to the above systematic limit set by the instrument calibration \citep[e.g.][]{Nevalainen17, Rasmussen}. Thus, in this work we adopt $\log N_{\ion{O}{VII}}($cm$^{-2})$ = 15 as the current X-ray detection limit for \ion{O}{VII}.

Since most of the intergalactic oxygen ($\sim$ 76\%; see Sect. \ref{Filaments}) lies within filaments, we estimate the depth through the absorbing material to vary between the order of $\sim 1$ Mpc (a conservative scenario) and $\sim 10$ Mpc (an optimistic scenario) for a filament crossing perpendicular to or along the line of sight, respectively.  Thus, we assume that the number density range between $\log n_{\ion{O}{VII}}($cm$^{-3}) = -9$ (for the path of $\sim 1$ Mpc) and $\log n_{\ion{O}{VII}}($cm$^{-3}) = -10$ (for a path of $\sim 10$ Mpc) represents the detectable \ion{O}{VII}. In addition, we want to investigate the metal enrichment of the large-scale structure, independently of the ionisation processes.  We do this by analysing the oxygen number density distribution above the same numerical range as in the case of \ion{O}{VII}. This approach effectively assumes that all oxygen denser than the above cut is ionised to \ion{O}{VII}. While this assumption is clearly too optimistic, it provides an approximate way of investigating the large-scale distribution of oxygen related to the observable \ion{O}{VII}. We do not utilise this assumption when estimating the actual \ion{O}{VII} detection probabilities.
 
\subsection{Column densities}
\label{column_density}
While the number density describes the physical, three-dimensional distribution of the studied ions, what is actually observed in the plane of sky is the two-dimensional projection of these structures. In our case, this corresponds to the integrated ion number densities along the line of sight observed as \ion{O}{VII} column densities. 
Thus, in order to generate predictions for future observations we produced a set of column densities from the ion number densities in EAGLE. 

For this we followed the method described in \citet{2019MNRAS.488.2947W}. We first divided the simulation box into a set of 20 Mpc long slices along one axis (see Sect. \ref{Detection} for the motivation for this depth), and subsequently divided each slice into squares of equal width and height along the remaining axes. 
We settled on a square size of  31.25$^2$ kpc$^{2}$, which is small enough so that results have converged \citep[see Appendix A in][]{2019MNRAS.488.2947W} and large enough to have adequate numbers of particles for robust statistics.
The width and height of the squares being much smaller than the depth of the slices, resulted in a number of thin, elongated segments or `columns'. 

The next step was to calculate the \ion{O}{VII} densities within each segment. To this end, we had to account for the nature of the SPH particles: each particle represents a volume in space, given by its smoothing length. Thus, we used a smoothing function (i.e. kernel) to distribute the number of ions within each SPH particle to the segment the particle is in and the surrounding segments that are covered by its kernel. Since the end result is a two-dimensional column density map, this smoothing was done only along the plane of the map. Along the direction of the projection we simply positioned all particles in the centre of each segment. Following \citet{2019MNRAS.488.2947W}, we used a C2-kernel \citep{wendland1995}, as the same kernel was used for the hydrodynamic properties of the simulated particles in EAGLE. After repeating the process for all the SPH particles and distributing the \ion{O}{VII} ions into the corresponding segments, the computed number of ions was divided by the aforementioned area to obtain the column density within each projected segment. Finally, in order to maximise the information from the relatively small EAGLE box, we performed the integration in all 3 orthogonal directions.

\subsection{Filaments}
\label{filaments}

In this work we utilised the results of the Bisous filament detection method  \citep{2007JRSSC..56....1S,2010A&A...510A..38S,2014MNRAS.438.3465T,2016A&C....16...17T} applied to the EAGLE galaxy catalogues of \citet{mcalpine_helly_etal_2016} as published in \citet{2020A&A...639A..71K} and \citet{2021A&A...646A.156T}.
In brief, the Bisous method uses the spatial distribution of galaxies as tracers of the Cosmic Web. It fits cylinders to galaxy overdensities, and subsequently connects and aligns said cylinders. 
%The process is repeated 1000 times, with regions more often covered by the cylinders being identified as filaments. 
The process is repeated 1000 times to account for the variation due to the stochastic nature of the Bisous method. The regions covered by the cylinders more frequently than an empirically defined critical rate are identified as filaments \citep{2014MNRAS.438.3465T}. 
The Bisous method delivers the volumes and the central spines of the filaments.
The current Bisous set-up is fine-tuned for detecting filaments at radial scales of $\sim$ 1 Mpc (see \citet{2021A&A...646A.156T} for details for testing this with EAGLE).

In order to compare the filaments recreated in EAGLE with the filament catalogue by \citet{2014MNRAS.438.3465T}, constructed using galaxies in the Sloan Digital Sky Survey \citep[SDSS;][]{2014ApJS..211...17A}, a magnitude cut (identical to that in \citet{2020A&A...639A..71K} and \citet{2021A&A...646A.156T}) was applied. Namely, only galaxies brighter than M$_{r} < -18.4$ were selected for constructing the Bisous filaments in EAGLE. Essentially, the applied magnitude cut sets a limit for the number density of galaxies (the key property employed by Bisous), and consequently determines the set of detected filaments \citep{Muru:2021}. By ensuring that the number density of the galaxies used by Bisous was the same for EAGLE and SDSS, we were able to extract a sample of EAGLE filaments comparable to observations.

To test the reliability of the results given by Bisous, \citet{2021A&A...646A.156T} studied how Bisous compares to a very different filament finding method, MMF/NEXUS+ \citep{2007A&A...474..315A, 2013MNRAS.429.1286C}. While Bisous uses galaxies as points to trace the filaments, NEXUS+ employs the geometry of a density field (dark matter density for the comparison with Bisous). Despite the very different methods, the thermodynamic properties of the WHIM in the respective filaments agreed within 10\%. In a similar vein, \citet{2019MNRAS.487.1607G} compared galaxy properties within Bisous and NEXUS+ filaments using the EAGLE simulation. They found a large overlap in the filamentary galaxy populations, similar fraction of stellar mass within filaments (a difference smaller than 10\%) as well as good alignment in the orientation of the filaments detected by the two methods.

However, regardless of the filament finding method, there are great variations in the temperatures and densities  between individual filaments detected with the same method \citep[see e.g.][]{2021A&A...649A.117G}. Thus, in order to select the most promising filaments to detect the missing baryons, we used the galaxy luminosity density \citep[LD; see][]{2012A&A...539A..80L} to divide the filaments into low, medium and high LD groups as in \citet{2021A&A...646A.156T}. As shown in \citet{2015A&A...583A.142N} and \citet{2022MNRAS.513.3387H}, the WHIM density and galaxy LD are tightly correlated. Moreover, the advantage of using the galaxy luminosity to classify filaments comes from its applicability to observations. The LD field was constructed by smoothing each galaxy's luminosity three-dimensionally and evaluating the field at discrete locations, such as the filament spines. The filaments were then divided based on an empirical classification into the aforementioned low, medium and high LD groups.  As shown in \citet{2021A&A...646A.156T}, selecting the filaments within highest LD regions ($\sim$10\% of the filaments of the full sample) yields the highest fraction of missing baryons. 
In the following analysis we used the total filament sample and, when so indicated, the high LD filament group to estimate the benefit of targeting the most WHIM-rich filaments.

\section{Oxygen enrichment of the cosmic filaments}
\label{Oxygen_analysis}
We first investigated the distribution of the intergalactic oxygen in the Cosmic Web filaments in the EAGLE simulation. This was done in order to gain an understanding of the metal enrichment of the large-scale structure in the low-redshift universe.
We study the ionisation of oxygen separately in Sect. \ref{ovii_3d_analysis}. As metals are created in stars within galaxies, from where a fraction of them are ejected towards the IGM, we first characterised their radial extent around haloes. We focused on the oxygen number densities outside the dark matter haloes defined by the FoF algorithm \citep{mcalpine_helly_etal_2016}. 
Subsequently we studied the volume and mass filling fractions of the intergalactic oxygen within the filaments.

\subsection{Intergalactic oxygen density profiles around galaxies}
\label{halo_profiles}
 \citet[][see their Fig. 2]{2020MNRAS.498..574W} showed that a significant fraction ($\approx$ 40\%) of the oxygen mass within the whole EAGLE simulation resides outside haloes. %defined via the application of the FoF algorithm on the dark matter particles (McAlpine...). 
 In order to quantify in detail the extent of the oxygen distribution in the IGM around haloes, we proceeded to construct radial profiles of the oxygen density centred at the central galaxy of a given halo\footnote{In most of the cases, the central galaxy coincides with the halo centre.}, %approximated with the halo center}
 defined via the subfind algorithm \citep{2001MNRAS.328..726S, 2009MNRAS.399..497D}, and catalogued by \citet{mcalpine_helly_etal_2016}. 

Since we ultimately aim to find the optimal $\ion{O}{VII}$ locations for X-ray detection, 
we selected haloes within the mass range associated with the highest $\ion{O}{VII}$ densities. \citet{2020MNRAS.498..574W} showed that the least massive haloes ($\sim \log M_{200}(M_{\odot}) < 11.5$) have too low virial temperatures compared to the $\ion{O}{VII}$ ionisation temperature in order to significantly contribute to the $\ion{O}{VII}$ budget. Similarly, haloes with $\log M_{200}(M_{\odot}) > 13.5$ have too high temperatures and most of the $\ion{O}{VII}$ has ionised out.
Thus, for the profile we selected haloes within the mass range of $\log M_{200}(M_{\odot}) = 12-13.5$.
In the simulation there is a total of 1282 haloes within this mass range ($\approx 5\%$ of all haloes), with virial radii between 210 and 660~kpc.
 
We first assigned each non-halo particle to the halo whose centre was closest. We then divided the volume around each mass-selected halo into spherical shells of increasing width and computed the number of non-halo particles in these shells. 
 We used the density profiles around the individual haloes to accumulate a distribution of density values at each given radial shell. We adopted the medians of the distributions in each shell as the representative values for the final oxygen density number profile.

The resulting oxygen density profile decreases rapidly with larger radii (see Fig. \ref{haloprofiles_plot}).
At a distance of $r \approx 1$ Mpc, the median oxygen density falls below $\log n_{O}($cm$^{-3}) \approx -10$. This corresponds to our optimistic observational limit considering an ideal scenario for X-ray observations whereby all of the oxygen in a given filament has been ionised to $\ion{O}{VII}$, and said filament is aligned with the line of sight such that the ion absorption path is 10 Mpc (see Sect. \ref{ejected}). In a more realistic scenario of 1 Mpc path through the filament, the median density drops below the detection limit ($\log n_{O}($cm$^{-3}) \approx -9$) at a distance of r $\approx$ 500 kpc, corresponding to $\sim 1-2 \times R_{200}$ in our mass range. The scatter of the densities at a given radius is large due to the wide range of halo masses considered. However, the extent of the detectable oxygen does not exceed 1 Mpc when focusing on narrower mass bins.

\subsection{Intergalactic oxygen within filaments}
\label{Filaments}
Having characterised the radial intergalactic oxygen density distributions around the central galaxies of the dark matter haloes, our next step was to utilise the knowledge of the distribution of the galaxies in the Cosmic Web 
%{\bf (extensively studied by e.g. Tartu group, talk with Elmo and Pekka)} 
in order to understand the metal enrichment in the filaments. \citet{2019MNRAS.487.1607G} showed that in our chosen mass range of $\log M_{200}(M_{\odot}) = 12-13.5$ (see above), the majority of the haloes in EAGLE are located within the cosmic filaments \citep[see also][for a different simulation]{2013MNRAS.429.1286C,2014MNRAS.441.2923C}. Similarly, 76\% of the intergalactic oxygen mass in EAGLE is located within Bisous filaments. Thus, if haloes are evenly spaced and are not located too far from one another, it would be reasonable to expect the oxygen to be uniformly distributed within filaments (as regularly done when using X-ray absorption measurements to estimate the baryon content of the filaments).%For halo masses of $\log M(M_{\odot}) = 12-13.5$, ,

In order to estimate the average separation between haloes within filaments, we first computed their number density in EAGLE. For haloes within the mass range of $\log M_{200}(M_{\odot}) = 12-13.5$ and closer than 1 Mpc to any Bisous filament spine, the resulting halo density is $\approx 0.05$ Mpc$^{-3}$. 
Assuming that they were homogeneously distributed throughout the filaments (i.e. each halo occupying a cube of similar size), the mean distance between haloes would be $\approx$ 3 Mpc. In this case, the galaxies would be too widely separated to efficiently enrich the filaments with oxygen. %$\ion{O}{VII}$. }

However, in reality the spatial galaxy distribution is far from homogeneous. Namely, along with dark matter and WHIM density, the galaxy density increases 
from the edge of a filament towards the spine. Thus, to estimate the expected shorter mean galaxy distance at the cores of the filaments due to enhanced mean density, we first computed the number density of central galaxies in 
the low end of 
our adopted mass range ($\log M_{200}(M_{\odot}) = 12.0-12.5$) of EAGLE haloes within a 0.2 Mpc radius of Bisous filament spines. The resulting density was $\approx 0.4$ Mpc$^{-3}$, 
enhanced by a factor of $\sim$ 10 from that within the approximate radius of a filament, r = 1 Mpc \citep[see][for similar values in a different simulation]{2020A&A...641A.173G}. This implies a mean distance between low mass haloes of $\approx$ 1 Mpc at the core regions of the filaments.
This is comparable to the radial extent of the oxygen density above the approximate detection limits. Thus, galaxies may be packed densely enough only at the very core regions of the filaments to efficiently enrich the intergalactic space with oxygen %OVII
at the detectable levels.

Another obvious deviation of the homogeneous spatial galaxy distribution is due to galaxies clumping into groups. It is a well established observational fact that the larger the galaxy mass and higher its luminosity, the more likely it is found in a group \citep{2009A&A...495...37T}. On the other hand, a significant contribution of intergalactic oxygen comes from the high-mass end of our sample. This suggests a clumpy oxygen distribution around galaxy groups. Thus, one would expect spatial segregation: low mass galaxies filling larger filament volumes with oxygen at the core regions of filaments while the high mass galaxies add a patchier component (see the next section for the volume filling fractions).

\subsubsection{Oxygen volume fractions and mass distributions}
\label{volume_filling}

We computed the volume filling fraction, that is, how large a fraction of volume within filaments is filled by oxygen above a given density level. To this end, we used the volume covered by individual particles calculated from their SPH variables, by dividing the mass of each simulated particle by its density\footnote{Within Bisous filaments, the sum of the volumes of individual particles equals the total volume covered by the filaments ($\approx 5\%$ of the total simulation volume).}. Then, the particle volumes above a given density were co-added to obtain a cumulative filament volume filling fraction.

We found that when considering the total Bisous filament volumes, (i.e. also including volumes within haloes), $\approx 11\%$ is filled with oxygen (see Fig. \ref{o_ovii_volume}) at densities above $\log n_{O}($cm$^{-3}) = -10$ (i.e. at approximately the currently detectable level in optimal conditions; see Sect. \ref{ejected}), while only 1\% above the more realistic limit $\log n_{O}($cm$^{-3}) = -9$. Thus, the worrying indication seen in EAGLE is that most of the Cosmic Web filament volume appears to contain oxygen below the current detection threshold, with oxygen above the limit forming relatively narrow envelopes around the haloes.

\begin{table}
\caption{Volume filling fractions and mass fractions of the intergalactic oxygen and $\ion{O}{VII}$ within Bisous filaments above the detectable densities. All masses are given as fractions of the total mass within the simulation.
    }         % title of Table
\label{table:volume_mass}      % is used to refer this table in the text
\centering                          % used for centering table
\begin{tabular}{c c c}        % centered columns (4 columns)
\hline\hline                 % inserts double horizontal lines
     & $\log n($cm$^{-3}) > -9$ & $\log n($cm$^{-3}) > -10$ \\    % table heading 
\hline                        % inserts single horizontal line
O volume   & 1\% & 11\% \\      % inserting body of the table
O mass     & 20\% & 35\% \\
\hline
$\ion{O}{VII}$ volume  & 0.4\% &  4\%   \\
$\ion{O}{VII}$ mass  & 20\% &  45\%   \\
%15.0 &  11\% (22\%) & 45\% (50\%)     \\
%15.5   & 1.5\% (4\%) & 8\% (10\%) \\
\hline                                   %inserts single line
\end{tabular}
\end{table}

%-----------------------------------------------------------------

   \begin{figure}
   \centering
   \includegraphics[width=\hsize]{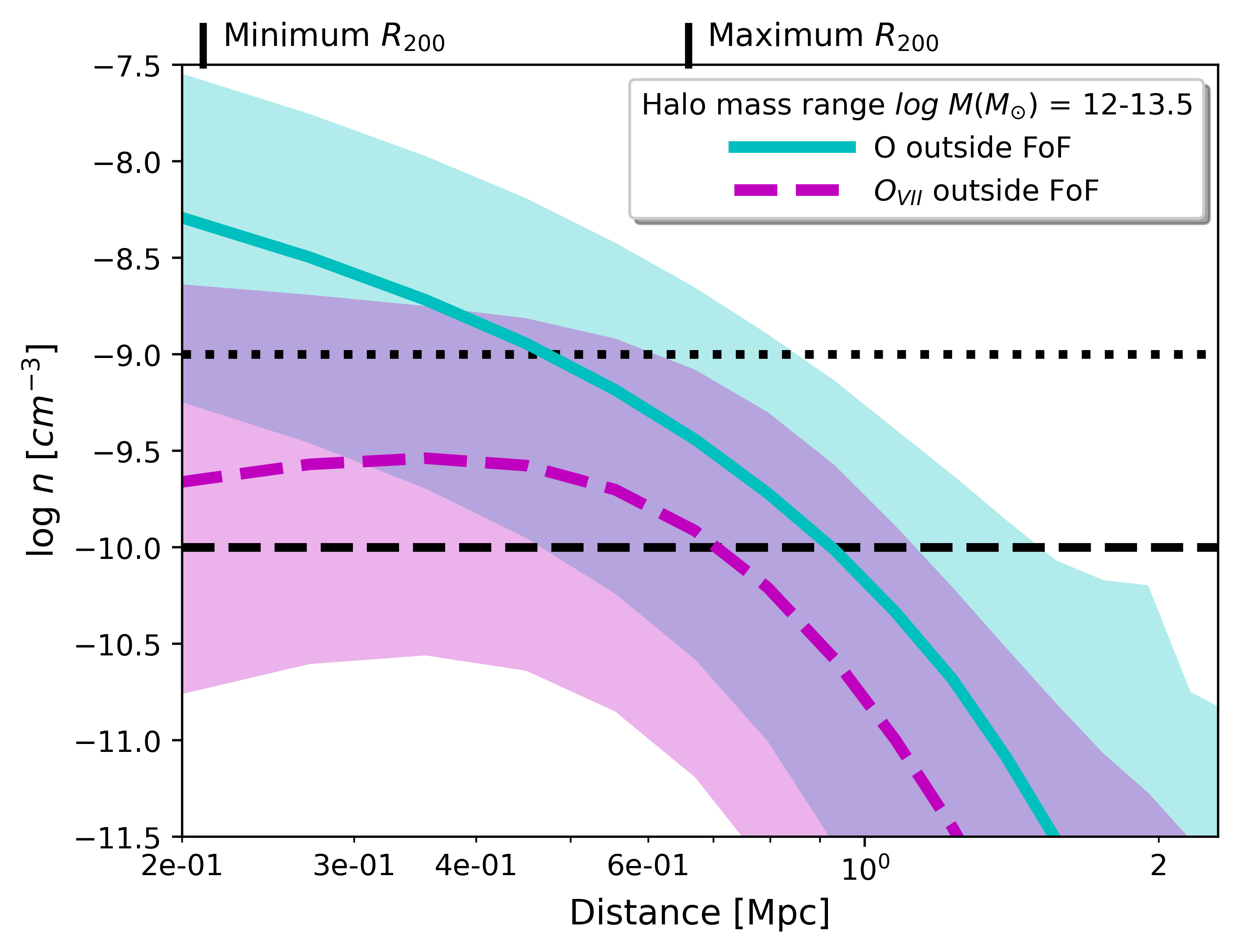}
      \caption{ Median number density (no weighting applied) of the intergalactic oxygen (turquoise line) and $\ion{O}{VII}$ (dashed purple line) as a function of distance from nearest halo within Bisous filaments. The shaded regions indicate the 68\% scatter in the density distribution from all the filaments combined. Within each distance bin, only particles not belonging to any FoF group were included. Haloes were selected within the mass range $\log M_{200}(M_{\odot})=$ 12 - 13.5. The corresponding range of $R_{200} = 220 - 660$ kpc is indicated on the upper axis. The horizontal dashed lines indicate the approximate detection limits with future X-ray detectors, with the lower and higher limits representing a filament parallel or perpendicular to the line of sight, respectively. 
              }
         \label{haloprofiles_plot}
   \end{figure}
%-----------------------------------------------------------------

   \begin{figure*}
    \begin{minipage}{0.48\textwidth}
    \centering   
    \includegraphics[width=\hsize]{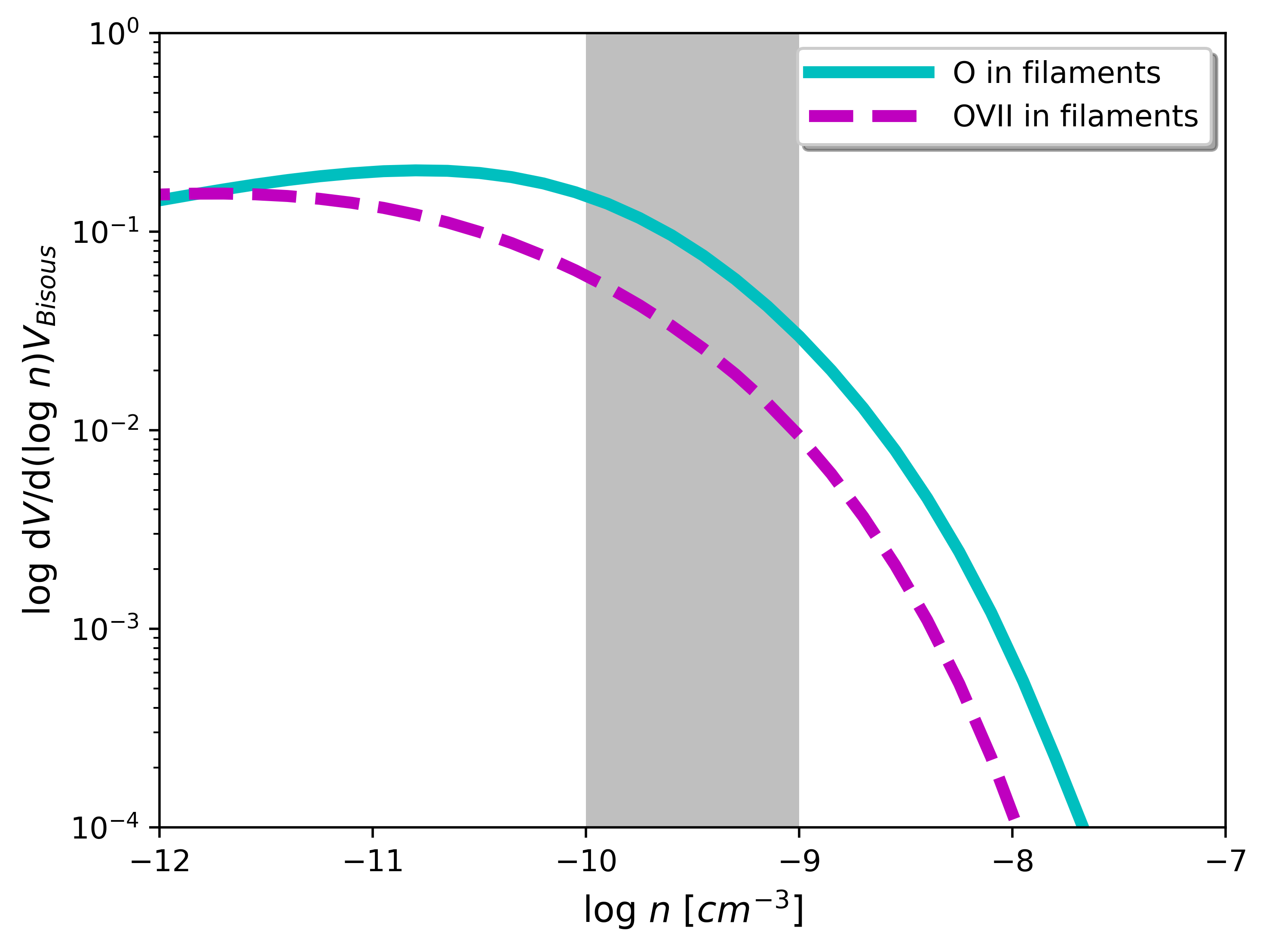}
    \end{minipage}
    \begin{minipage}{0.48\textwidth}
    \includegraphics[width=\hsize]{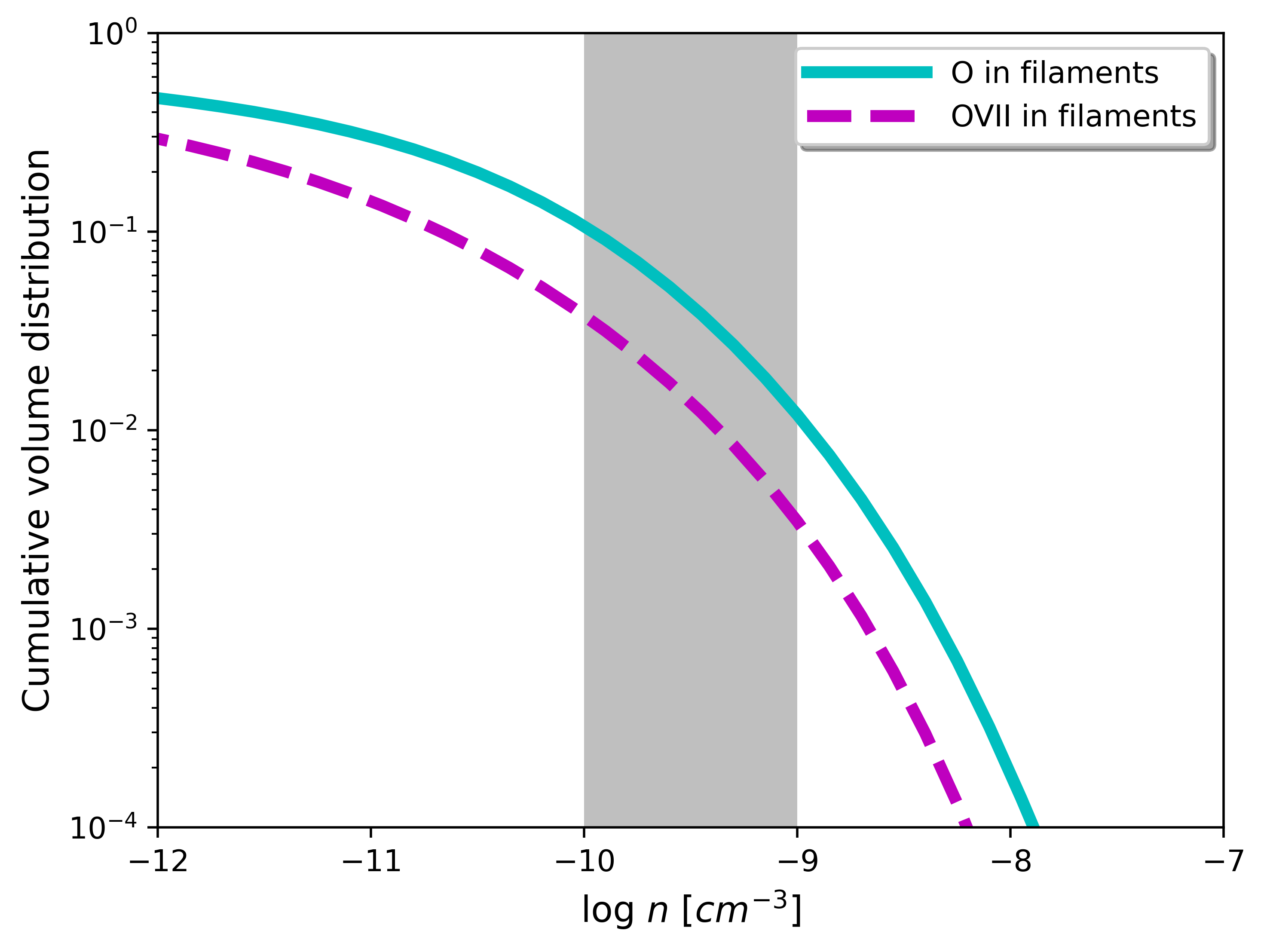}
    \end{minipage}
    \caption{Filament volume fractions filled by oxygen and \ion{O}{VII}. \textit{Left panel:} Volume filled by oxygen (turquoise line) and \ion{O}{VII} (dashed purple line) at a given number density within the IGM in Bisous filaments, as a fraction of total Bisous volume. \textit{Right panel: } Fraction of the volume above a given number density $n$ (cm$^{-3}$) for oxygen (turquoise line) and \ion{O}{VII} (dashed purple line) within Bisous outside haloes normalised by the total Bisous volume. The grey shaded area indicates the assumed observational limits, for a path between 1 Mpc (log $n \approx$ -9) and 10 Mpc (log $n \approx$ -10) along the line of sight.
              }
     \label{o_ovii_volume}
   \end{figure*}
%-----------------------------------------------------------------

   \begin{figure*}
    \begin{minipage}{0.48\textwidth}
    \centering   
    \includegraphics[width=\hsize]{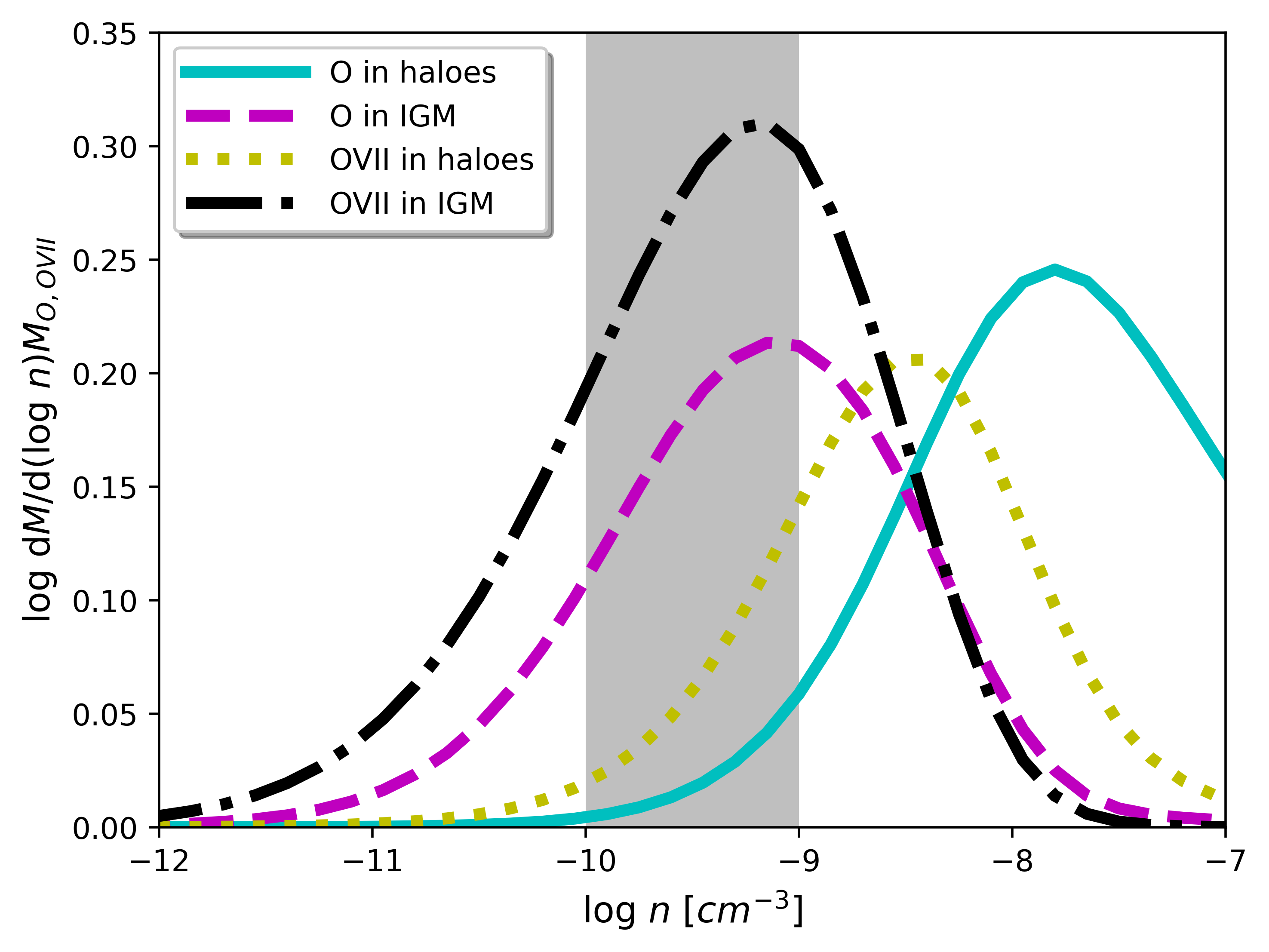}
    \end{minipage}
    \begin{minipage}{0.48\textwidth}
    \includegraphics[width=\hsize]{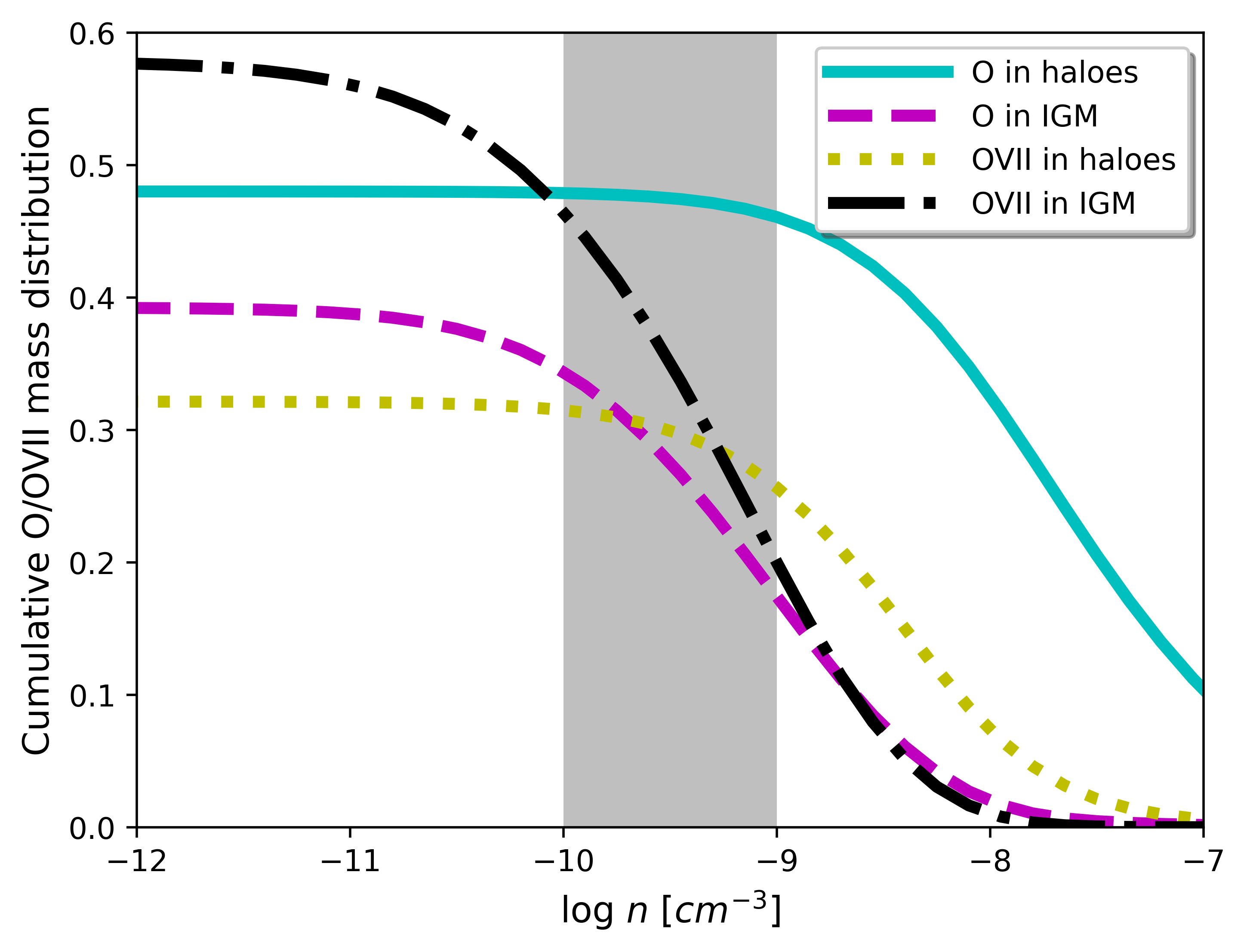}
    \end{minipage}
    \caption{Oxygen and \ion{O}{VII} mass fractions within filaments. \textit{Left panel:}  Distribution of filamentary oxygen and \ion{O}{VII} mass fractions in haloes and the IGM as a function of number density, divided by the total oxygen and $\ion{O}{VII}$ mass in the gas phase, respectively. \textit{Right panel: } Cumulative mass fraction within Bisous filaments. The sum of oxygen and \ion{O}{VII} mass above a given density was divided by the total (non-stellar) mass of oxygen and \ion{O}{VII} in the simulation, respectively. In both panels the solid turquoise (dotted yellow) and dashed purple  (dash-dotted black) lines refer to the oxygen (\ion{O}{VII}) mass within haloes and IGM in filaments, respectively. The grey shaded area indicates the assumed observational limits, for a path between 1 Mpc (log $n \approx$ -9) and 10 Mpc (log $n \approx$ -10) along the line of sight. In both panels we show only the oxygen and \ion{O}{VII} distributions within Bisous filaments. The remaining oxygen and \ion{O}{VII} are outside the filaments.  
              }
     \label{o_ovii_mass}
   \end{figure*}
%-----------------------------------------------------------------

We proceeded to better understand the implied low filament volume filling fraction of detectable oxygen.  
To this end, we computed the oxygen mass distribution as a function of density within filaments, separately for haloes and the IGM. 
In filamentary haloes (i.e. haloes within the Bisous filaments), the oxygen distribution peaks at a density of $\log n_{O}($cm$^{-3}) \approx -7.8$ (Fig. \ref{o_ovii_mass}, left panel). 
On the other hand, the intergalactic oxygen peaks at an order of magnitude lower density of $\log n_{O}($cm$^{-3}) \approx -9.1$. 
Integrating the distributions we found that a significant fraction (between 20--35 \%) of the total oxygen mass in the EAGLE simulations lies within the filamentary IGM above the approximate current X-ray detection limit (Fig. \ref{o_ovii_mass}, right panel). This corresponds to $\approx 50-85\%$ of the total intergalactic oxygen mass, which we found in Sect. \ref{volume_filling} to cover only 1-11\% of the filament volumes. Thus, the intergalactic oxygen peak probably corresponds to small and dense pockets around haloes (see Sect. \ref{ovii_distribution} for the radial distribution of $\ion{O}{VII}$).

\subsubsection{Oxygen density profiles around filament spines}
\label{spineprofile}
Since (1) the galaxy density increases when approaching the spine of the filament \citep[e.g.][]{2020A&A...641A.173G} and (2) the oxygen density at detectable levels is confined within 1 Mpc of the halo centre, %(see Section \ref{halo_profiles}), 
it is expected that the oxygen density increases towards the filament spines.
In order to quantify the expected radial dependence of the oxygen density
we proceeded to create radial oxygen density profiles as a function of distance from the filament spines. As our region of interest lies in the IGM, we selected only simulated particles outside FoF haloes. Analogously to the halo profiles in Sect. \ref{halo_profiles}, each particle was assigned to its closest filament spine from which the distance was calculated. The subsequent profiles were done in  a similar fashion to \citet{2021A&A...646A.156T} for the baryon density, that is, separating the volumes around filament spines into concentric hollow cylinders, and dividing the total number of oxygen atoms within each cylinder by its volume. In addition, we removed the volumes covered by the removed haloes from each concentric cylinder. After computing the radial profiles for individual filament spines, we selected the median density value at each distance bin, thus producing a single profile. This was done separately for filaments in different LD groups (see Sect. \ref{filaments} for the different LDs).
  
\begin{figure}
   \centering
   \includegraphics[width=\hsize]{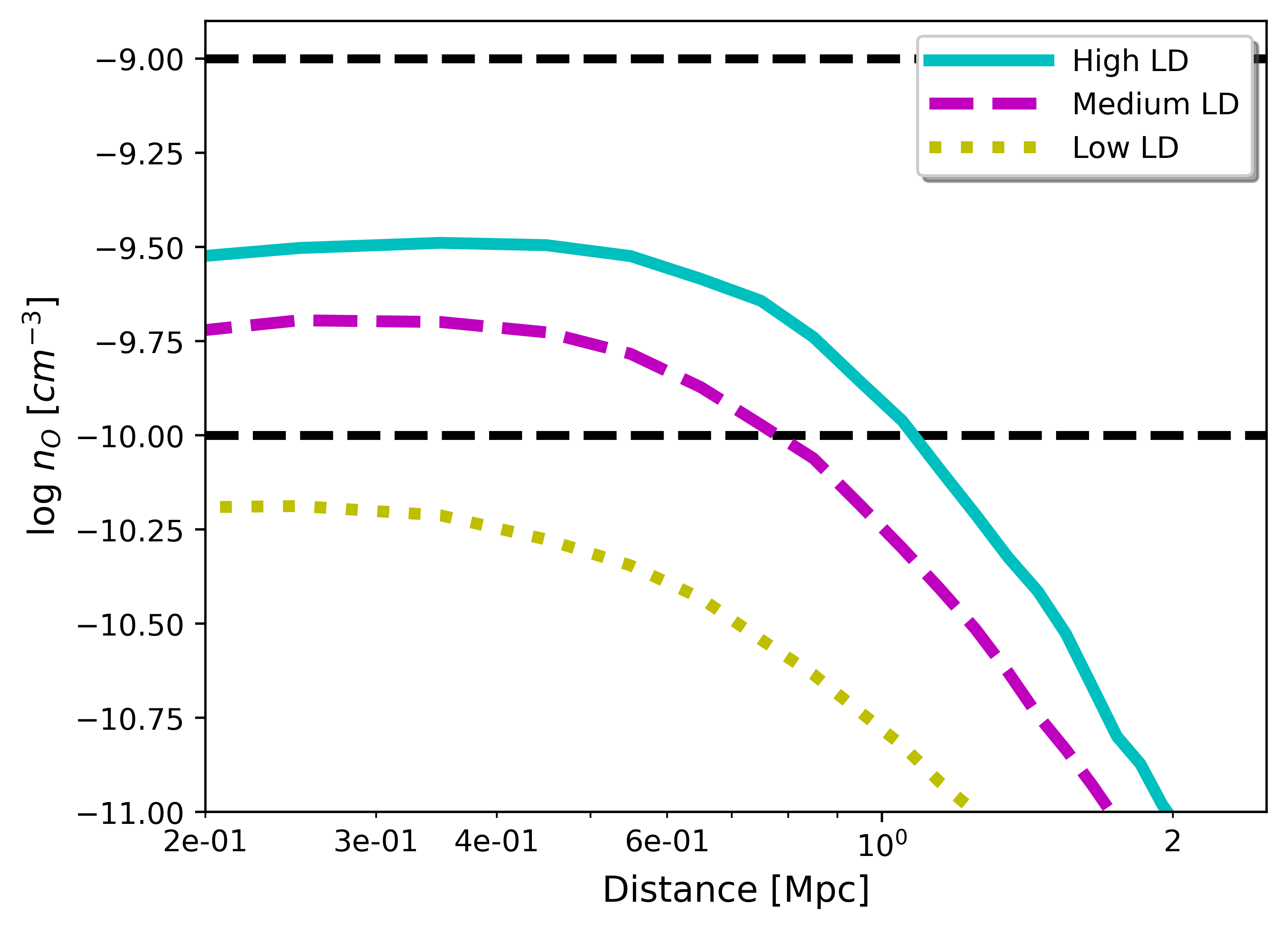}
      \caption{Number density ($n_{O}$) profiles of the intergalactic oxygen as a function of distance from filament spines. Each profile represents the median value from the individual spines at each given distance bin, with turquoise, dashed purple, and dotted yellow lines showing the median values for high, medium, and low LD filaments, respectively. The approximate detection limit range is shown with horizontal dashed lines.
              }
         \label{filamentprofiles_plot}
   \end{figure}

 Similarly to the halo profiles, the oxygen density profiles in filaments peak at the spine and decrease rapidly with radius (Fig. \ref{filamentprofiles_plot}). Moreover, high LD profiles yield oxygen densities $\approx$ 2 and $\approx$ 5 times greater than medium and low LD profiles, respectively. This is to be expected, as filaments within higher LD regions contain more galaxies and therefore more stars producing oxygen, which is then expelled into the IGM. The oxygen density for medium and high LD regions remains above the limit of $\log n_{O}($cm$^{-3}) = -10$ up to a distance of $r = 1-1.2$ Mpc from the spine. However, the profiles do not reach the more realistic limit of $\log n_{O}($cm$^{-3}) = -9$. This suggests that not even the high luminosity filaments, containing most of the hot WHIM \citep{2021A&A...646A.156T}, enclose enough oxygen to be detected perpendicular to the line of sight.

\begin{figure}
   \centering
   \includegraphics[width=\hsize]{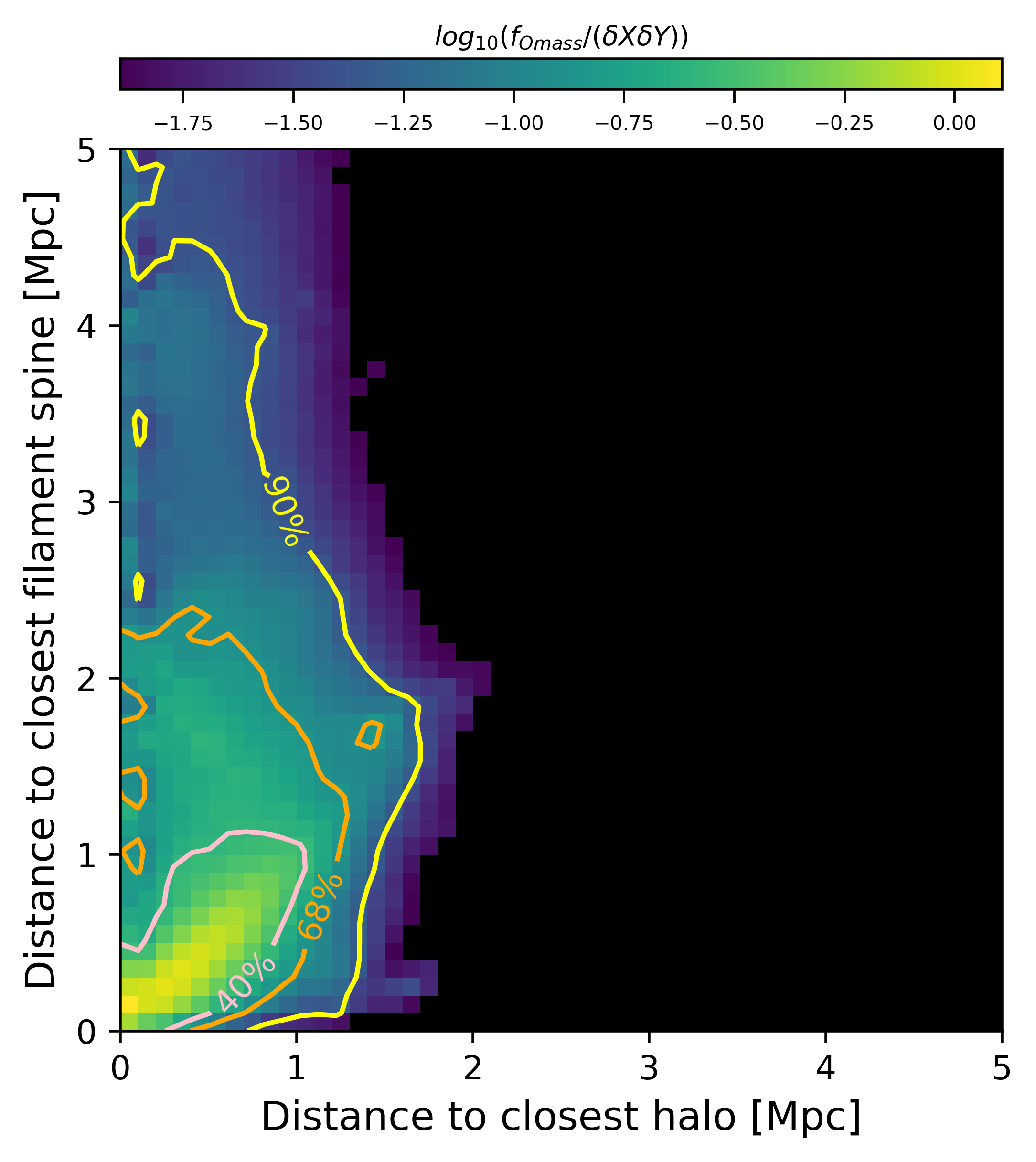}
      \caption{  Oxygen mass fraction towards halo centres %{\bf mass range...} 
      and filament spines. Each simulated particle was positioned within one pixel in the diagram based on the distance to its closest halo centre and filament spine. Subsequently, the oxygen mass contained in the particles was co-added within each pixel and divided by the pixel size. Finally, the value in each pixel was normalised with the highest value in the diagram. The colour map shows values down to 1\% of the peak value. The pink, orange, and yellow contour lines enclose 40\%, 68\%, and 90\% of the oxygen mass, respectively.  
              }
         \label{filament_halo_distance_plot}
   \end{figure}

We then proceeded to connect the oxygen density profiles around filament spines to those around haloes. %(see Section \ref{halo_profiles}). 
To this end, we computed a mass diagram that reveals the oxygen distribution with respect to the distance to both halo centres and filament spines (Fig. \ref{filament_halo_distance_plot}). Within $\approx 1$ Mpc from filament spines and haloes there is a correlation in the oxygen mass. This can be explained by the fact that haloes (from where oxygen is expelled) tend to reside along or nearby the filament spines. However, this correlation between spines and haloes disappears at larger distances. For haloes, only $\approx$ 4\% of the total oxygen mass lies beyond 1.5 Mpc, while oxygen can be found at larger distances from filaments. Indeed, $\approx 38\%$ of the oxygen mass can be found further than 1.5 Mpc from the filament spines. At the same time, within our chosen mass range of $\log M_{200}(M_{\odot}) = 12 - 13.5$, $\approx 40\%$ of haloes reside further than 1.5 Mpc from any filament spine. This indicates that oxygen is more tightly related to the haloes where it is formed, and filaments play only a secondary role in its distribution. This also raises the question of the environment where these non-filamentary haloes and oxygen reside. While potentially most are in the nodes of the Cosmic Web (not traced by the Bisous method), a significant fraction might be in low-density regions and voids. However, this is beyond the scope of this work, since our aim is to trace the hot IGM, contained mainly within filaments. Thus, we moved on to examine the ionisation of oxygen into \ion{O}{VII}, the dominant ion at the hot WHIM temperatures.

%-----------------------------------------------------------------

%-----------------------------------------------------------------

%-----------------------------------------------------------------

%-----------------------------------------------------------------

\section{Ionisation of the cosmic oxygen}
\label{ovii_3d_analysis}
Given the above characterisation of the intergalactic oxygen, we proceeded to studying the ionisation of the cosmic oxygen via $\ion{O}{VII}$ distributions and temperature analysis.

%-----------------------------------------------------------------

   \begin{figure*}
   \centering
   \includegraphics[width=\hsize]{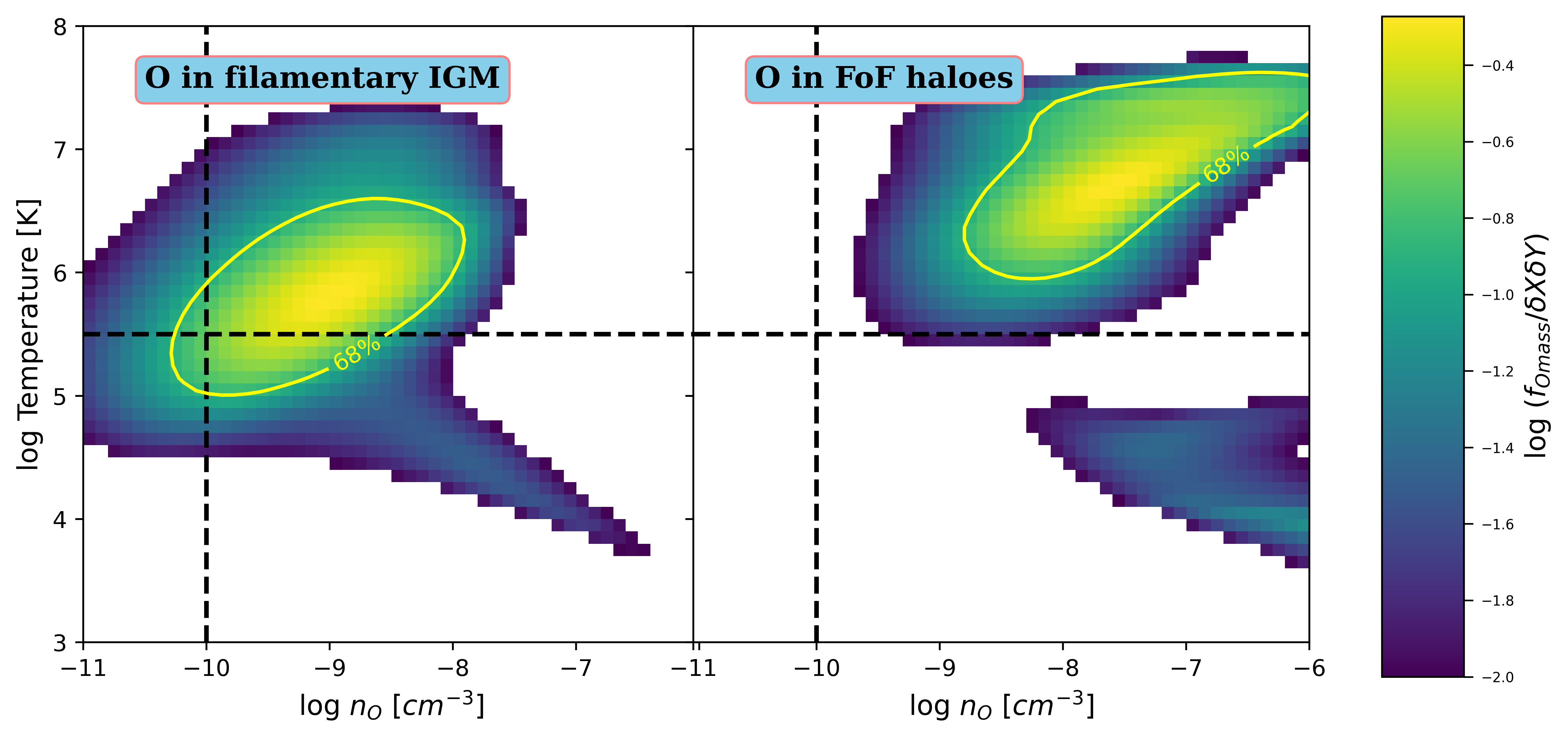}
   \includegraphics[width=\hsize]{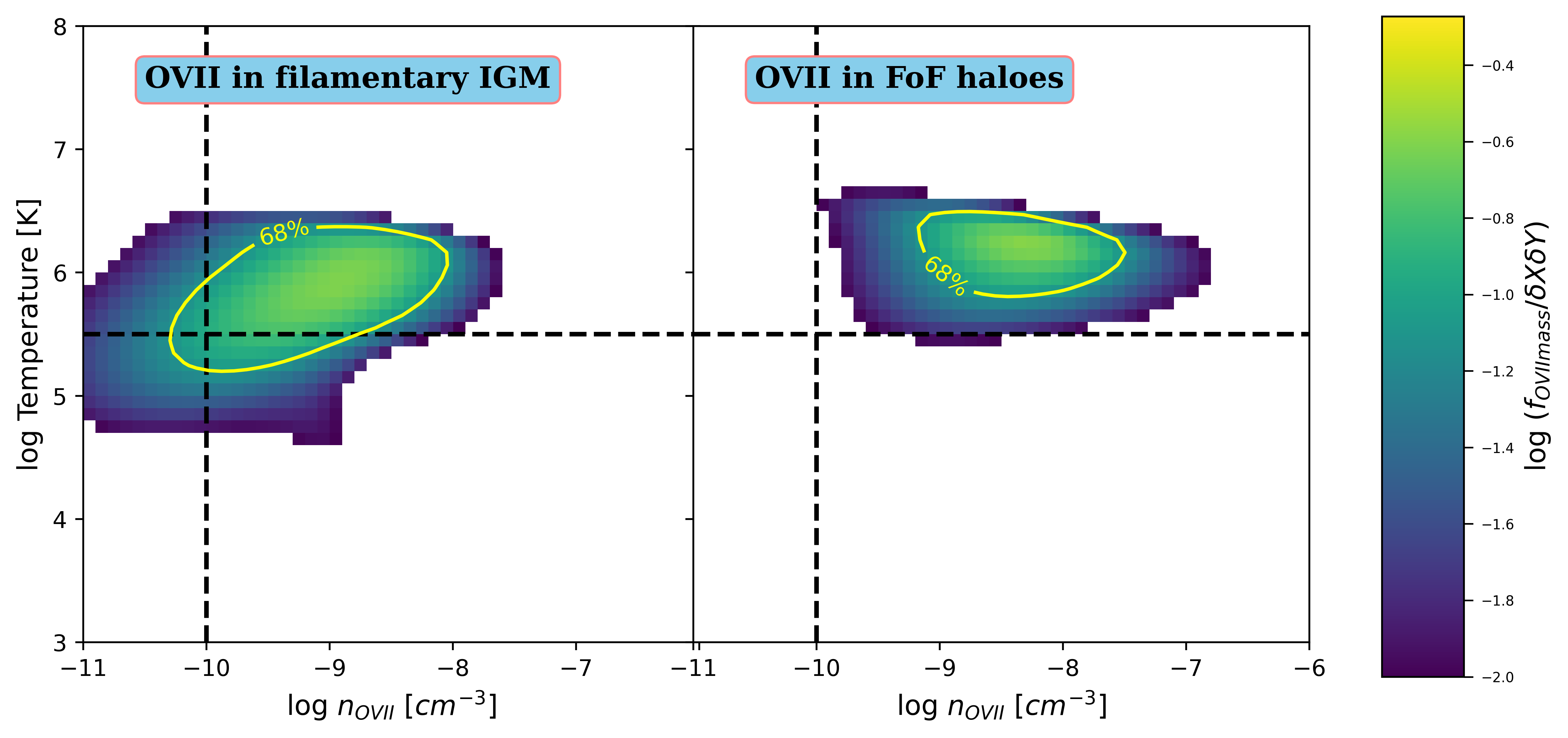}
      \caption{ Distribution of oxygen (upper panel) and \ion{O}{VII} (lower panel) mass as a function of number density and temperature. The left panels show the oxygen and \ion{O}{VII} distributions in the IGM within Bisous filaments. Gas in FoF haloes is shown in the right panels, where we selected haloes located within Bisous filaments. The colour scales indicate the fractions of oxygen and \ion{O}{VII} mass over the total oxygen mass in the same environment (divided by pixel area; values are shown down to 1\%). The yellow circles mark the regions where 68\% of the given mass is contained. The horizontal and vertical dashed lines mark the limit for the hot gas ($\log T(K) > 5.5 $) and the optimistic lower density limit for X-ray observations (given a 10 Mpc long path aligned with the line of sight), respectively. 
              }
         \label{o_ovii_phasediag}
   \end{figure*}
%-----------------------------------------------------------------

\subsection{Thermodynamic large-scale structure}
\label{heating}
Theoretically, the heating mechanisms of the intergalactic baryon gas in filaments are well established. The most important  process in this context is the shock-heating arising from the accretion of the baryons towards the filaments pulled by the dark matter gravity \citep[e.g.][]{Ryu2003}. The baryons falling towards the filaments also experience adiabatic compressional heating. Both effects are stronger closer to the maxima of the gravitational potential, that is, the filament spines. This region is thus expected to harbour the hottest WHIM,
as consistently demonstrated by numerous cosmological simulations
 \citep[e.g.][]{CenOst99,Dave01,Dolag06,2009ApJ...697..328B,Cui12,Cui18,2021A&A...646A.156T,Haider16}. 
The SN and AGN feedback as well as halo mergers provide additional heating in the close neighbourhood of galaxies. 

To give a quantitative example, \citet{2021A&A...646A.156T} showed that the density and the temperature of the intergalactic gas in the EAGLE simulations peak at the filament spines. 
When focusing on the inner $\sim 1$ Mpc regions around the densest filament sample, the temperatures reach a level of $\log T(K) \approx 5.5 - 6.0$ (i.e. the energies of most of the free electrons exceed the ionisation energy of $\ion{O}{VII}$).
The baryon overdensities in these regions reach the level of $\sim$ 10. 
In these conditions \ion{O}{VII} is mainly produced via collisional ionisation \citep[e.g.][]{Strawn:22_tmp}.
The demonstration that the oxygen density also peaks in these regions (see Sect. \ref{spineprofile}) suggests that the optimal place to look for the missing baryons in the form of the hot WHIM in X-rays are the central regions of the densest filaments. 

We performed here a quantitative temperature analysis in order to understand better the above described scenario of the ionisation of the cosmic oxygen. Indeed, we found that the majority ($\approx 72\%$) of the oxygen mass considering the full intergalactic volumes within filaments has a temperature above $\log T(K) > 5.5$ (see Fig. \ref{o_ovii_phasediag}), required for \ion{O}{VII} production under collisional ionisation equilibrium (CIE). %.. tästä vain pieni osa on yli log T 6.4 jotta OVII ionisoituisi OVIII:ksi... 
Assuming CIE, almost all of the oxygen is expected to be ionised to \ion{O}{VII} within the temperature range of $\log T(K) = 5.5 - 6.2$, while \ion{O}{VIII} begins gaining ground at higher temperatures \citep{Mazzotta:1998}. We found that $\approx 50\%$ of the intergalactic EAGLE oxygen mass is in this temperature range. However, %Since 
the intergalactic \ion{O}{VII} mass fraction is smaller ($\approx 33\%$), % it is 
implying that the ionisation responsible for the \ion{O}{VII} population is not purely CIE from a single phase gas.

The very low density of the filamentary IGM (density contrast $\delta \sim 1-100$) is probably responsible for the relatively low efficiency of the \ion{O}{VII} production. 
The efficiency of the collisional ionisation %ionsation effect of the CIE process
decreases relatively fast with lower densities, as it is proportional to the density squared. %(the rate for collisional ionisation is proportional to the density squared, n$^2$). 
Consequently, the collisional ionisation timescale is comparable to the Hubble time at a density contrast $\delta \sim 10$ and exceeds it significantly %exponentially?
at lower densities \citep[e.g.][]{Bykov:2008SSRv}. On the other hand, photoionisation begins to play a more significant role at lower densities due to its linear dependence on the ion density. 
Towards lower densities, the UV background photon density per particle increases and thus 
a larger fraction of \ion{O}{VII} is produced via photoionisation \citep[e.g.][]{2018MNRAS.477.450N,2019MNRAS.488.2947W,Strawn:22_tmp}. However, only the high-energy tail of the UV background photons exceeds the ionisation energy of \ion{O}{VII} \citep{2001cghr.confE..64H}, and therefore this effect does not compensate for the inefficiency of CIE in the production of \ion{O}{VII}. Additional factors, such as further photoionisation 
by the X-ray background, %and the low recombination rate in the IGM (proportional to density squared) 
may also play a role, but they are beyond the scope of this work.

Conversely, the bulk of the oxygen in the FoF haloes is hotter: most of the oxygen mass is at $\log T(K) > 6$ (see the upper-right panel in Fig. \ref{o_ovii_phasediag}).
Due to the very high temperatures, most of the oxygen is expected to be further ionised beyond \ion{O}{VII}.
Indeed, the fraction of oxygen in the form of \ion{O}{VII} in haloes is $\approx 12\%$, much smaller than in the intergalactic space.

\subsection{Intergalactic $\ion{O}{VII}$ around galaxies}
\label{ovii_distribution}
While the core regions of filaments have ideal temperatures for $\ion{O}{VII}$, %However, as already discussed, 
according to EAGLE the oxygen density distribution above the level approximately achievable with current and near-future X-ray detectors
% (log $n_{O}$ = {\bf -10 - -9}) 
is highly inhomogeneous (see Sect. \ref{halo_profiles}). %{\bf (even worse at 10-9...)} , 
It peaks at the locations of the haloes and drops rapidly below the detection level as a function of distance from the halo.

\begin{figure*}
    
    \includegraphics[width=\hsize]{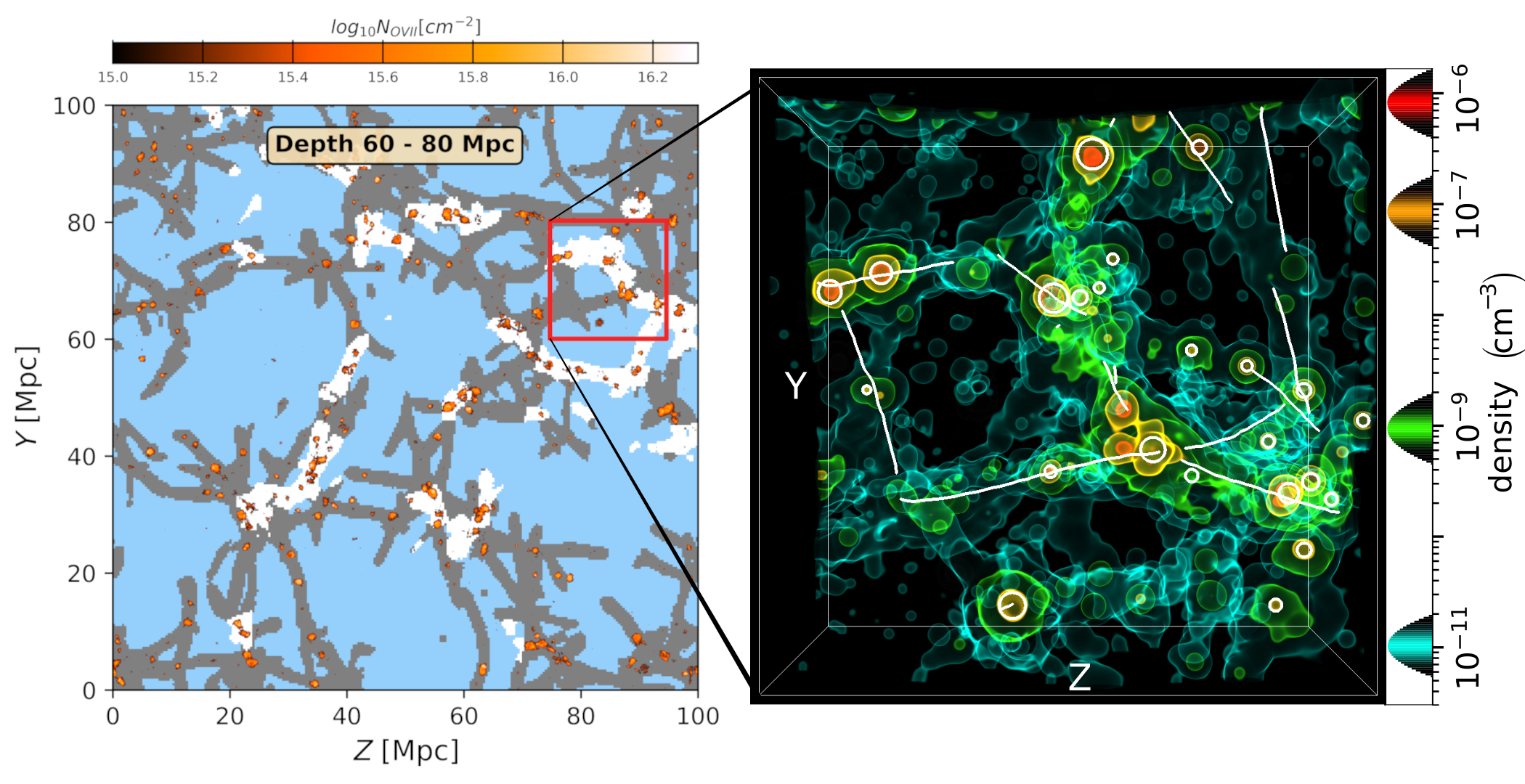}
      \caption{Comparison between the two-dimensional column densities and the three-dimensional number densities of \ion{O}{VII}. \textit{Left:} Column densities of \ion{O}{VII} projected through a 20 Mpc deep slice along the x-axis. The colour scheme indicates the \ion{O}{VII} column density with a lower limit at $\log N_{\ion{O}{VII}} ($cm$^{-2}) = 15$, while the Bisous filaments and high LD Bisous filaments are shown by grey and white areas, respectively.  This image is the same as the middle-right panel of the column density maps in Fig. \ref{phase_slices.fig}. The red square corresponds to the volume shown in the right-hand figure. \textit{Right:} Three-dimensional visualisation of the \ion{O}{VII} distribution in a (20 Mpc)$^{3}$ zoom-in box within the larger EAGLE volume. The coloured regions represent the \ion{O}{VII} density contours at given values, as indicated on the right vertical axis.
      The green contour shows the spatial extent of \ion{O}{VII} above the optimistic detection limit of $n_{\ion{O}{VII}} = 10^{-9}$ cm$^{-3}$ discussed in the text. The white lines show the Bisous filament spines, while 
      the white circles indicate the volumes within $R_{200}$ around the central galaxies of the FoF haloes. A three-dimensional visualisation is available as an online movie in the YouTube channel of Tartu Observatory, https://youtu.be/878twggEFoY.  }
     
         \label{ovii_3d_plot}
   \end{figure*}

In order to quantify the radial behaviour of $\ion{O}{VII}$ around haloes, we constructed an $\ion{O}{VII}$ density profile following the method we applied to oxygen (i.e. choosing only particles that reside in the IGM around haloes within a mass range of $\log M_{200}(M_{\odot}) = 12-13.5$). %, see Section \ref{halo_profiles}). 
We found that beyond the virial radius $R_{200}$, the $\ion{O}{VII}$ density 
%peaks approximately at the virial radius $R_{200}$ {\bf kuinka korkea r200 piikki on verrattuna at r=0?} and subsequently 
drops rapidly below the optimistic observational limit of $\log n_{\ion{O}{VII}} < -10$ at a radial distance of $\approx$ 700 kpc
(see Fig. \ref{haloprofiles_plot}). Considering the conservative observational limit of $\log n_{\ion{O}{VII}} < -9$ the picture is even bleaker, as the median density profile does not reach that level at any radii.  The visual investigation of the three-dimensional distribution of \ion{O}{VII} density indeed confirms the relatively low radial extent of \ion{O}{VII} at detectable levels (see Fig. \ref{ovii_3d_plot}).

The analysis above indicates that the envelopes containing detectable amounts of  $\ion{O}{VII}$ are very tightly constrained around the haloes. 
A similar conclusion can be derived from a work on the high-resolution HESTIA %\LEt{ Consider defining.}
simulation \citep{2022MNRAS.512.3717D}. It is a simulation of the Local Group, with the Milky Way and Andromeda galaxies, indicating that the median \ion{O}{VII} column density profile drops below the $\log N_{\ion{O}{VII}} ($cm$^{-2}) = 15$ level beyond $\sim 3 \times R_{200}$ from the halo centre \citep[their Fig. 4]{2022MNRAS.512.3717D}. In fact, the EAGLE \ion{O}{VII} envelopes are narrower than those of oxygen (%Section \ref{halo_profiles} and 
Fig \ref{haloprofiles_plot}). %This implies that 
Indeed, not all oxygen is ionised into $\ion{O}{VII}$, but instead there is a spatial co-existence of different temperature phases as well as limiting effects on the ionisation efficiency. 

%-----------------------------------------------------------------

   %\end{figure}
%-----------------------------------------------------------------

\subsection{Intergalactic \ion{O}{VII} within filaments}

Given that the ionisation acts as an additional constraint and thus the median $\ion{O}{VII}$ density profile around haloes remains lower than the oxygen density at all radii, we repeated the analysis of mass distributions and volume filling fractions for $\ion{O}{VII}$ within filaments (see Sect. \ref{volume_filling} for the oxygen volume and mass fractions). Additionally, we computed the fraction of hot WHIM that could potentially be traced with \ion{O}{VII}.
%leading to a lower spread of $\ion{O}{VII}$ from haloes, we analysed below its volume filling fraction and mass distribution within filaments.}

\subsubsection{$\ion{O}{VII}$ volume fractions and mass distributions in filaments}
\label{ovii_fractions}

In a similar fashion to oxygen, the intergalactic $\ion{O}{VII}$ mass distribution peaks at $\log n_{\ion{O}{VII}}($cm$^{-3}) \sim -9$ (see the left panel in Fig. \ref{o_ovii_mass}), at the upper boundary of the estimated detection limit (see Sect. \ref{ejected}). Above this density limit, approximately 20\% of all the $\ion{O}{VII}$ mass in the whole simulation is contained within filaments (see Table \ref{table:volume_mass} and right panel in Fig. \ref{o_ovii_mass}). Moreover, $\approx 45\%$ of the $\ion{O}{VII}$ is within filaments above a density limit of $\log n_{\ion{O}{VII}} > -10$. This corresponds to $\approx 77\%$ of the total intergalactic \ion{O}{VII} (i.e. outside haloes; see Fig. \ref{o_ovii_phasediag}). However, considering that the median density drops rapidly as a function of distance from haloes, this $\ion{O}{VII}$ mass is contained within small pockets.

Indeed, only 4\% and 0.4\% of the full EAGLE filament volume is filled by \ion{O}{VII} with densities above $\log n_{\ion{O}{VII}} = -10$ and $-9$, respectively (see Fig. \ref{o_ovii_volume}, right panel). At the higher density limit, a path of the order of 1 Mpc is enough to result in detectable column densities, thus rendering the halo outskirts observable. Consequently, the filamentary \ion{O}{VII} density above the expected detection levels is very patchy, which is evident in the column density maps (see the right panel in Fig. \ref{ovii_lim1314} and all panels in Fig. \ref{phase_slices.fig}; see Sect. \ref{column_density} for the details of producing the maps). Furthermore, the three-dimensional distribution reveals how the filament volumes are largely devoid of \ion{O}{VII} above detectable levels (see Fig. \ref{ovii_3d_plot}).

Thus, assuming that EAGLE captures accurately the distribution of \ion{O}{VII} in the Cosmic Web filaments, the above results have important consequences for the missing baryon detection. Namely, the essential assumption when evaluating the cosmic baryon content based on X-ray absorption measurements is that \ion{O}{VII} is well mixed into the WHIM. This assumption is not valid above the current and expected near future detection limit of $\log N_{\ion{O}{VII}} ($cm$^{-2}) > 15$.  Consequently, the projected sky area of the volumes filled by \ion{O}{VII} is relatively small at the detectable level, rendering its observation  challenging (see Sect. \ref{Detection} for the feasibility study).

\subsubsection{Hot WHIM traced by \ion{O}{VII}}

\begin{figure}
   \centering
   \includegraphics[width=\hsize]{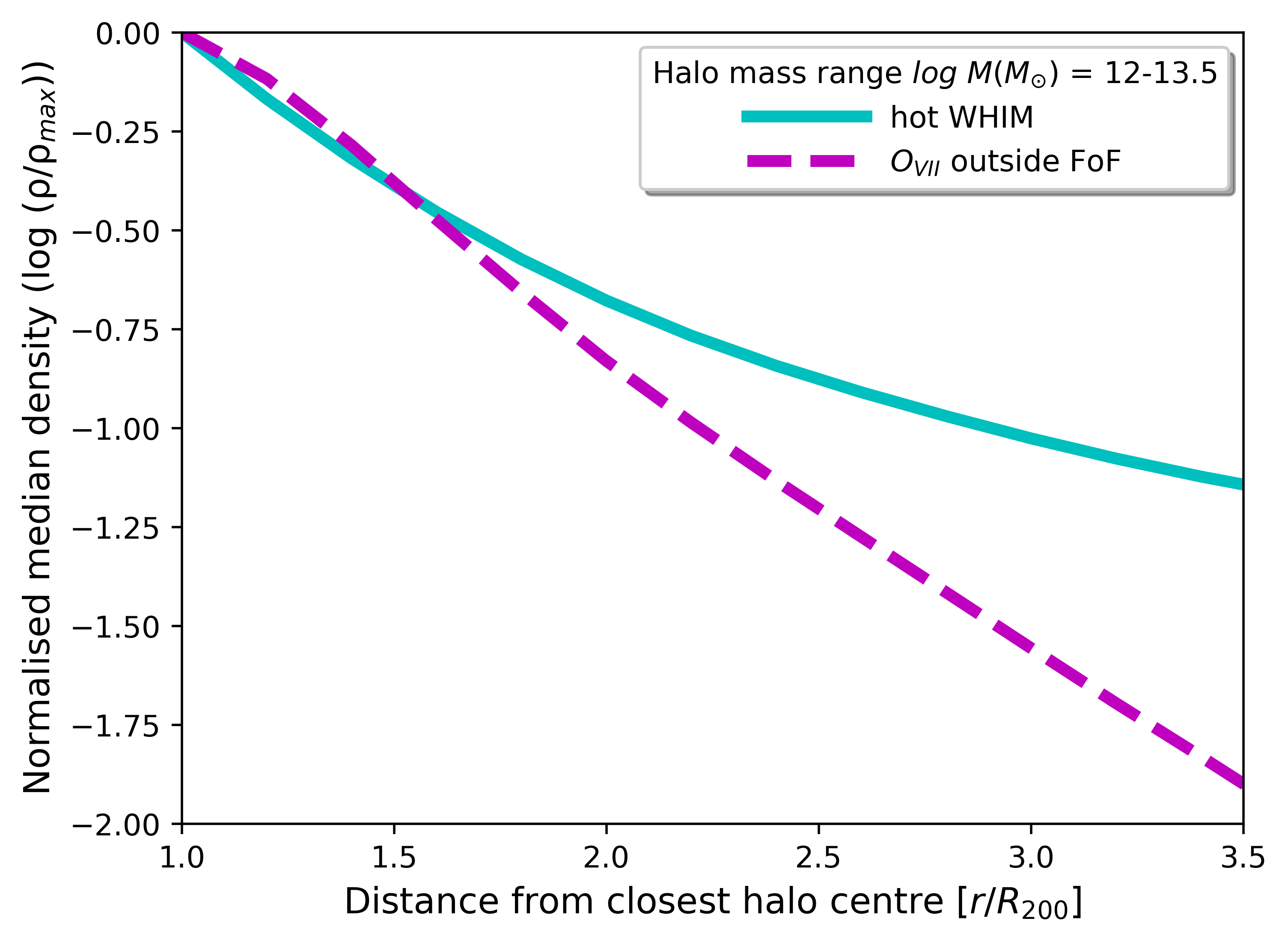}
      \caption{ Normalised density distribution of the hot WHIM (turquoise line) and $\ion{O}{VII}$ (dashed purple line) as a function of distance from the closest halo centre, in units of virial radius, $R_{200}$. %The shaded areas indicate the 68\% distribution around the median. 
      As in Fig. \ref{haloprofiles_plot}, haloes were selected within the range of $\log M_{200}(M_{\odot}) = 12 - 13.5$.}
        \label{hW_ovii_profiles.fig}
      \end{figure}

\begin{figure*}
    \begin{minipage}{0.31\textwidth}
    \centering   
    \includegraphics[width=\hsize]{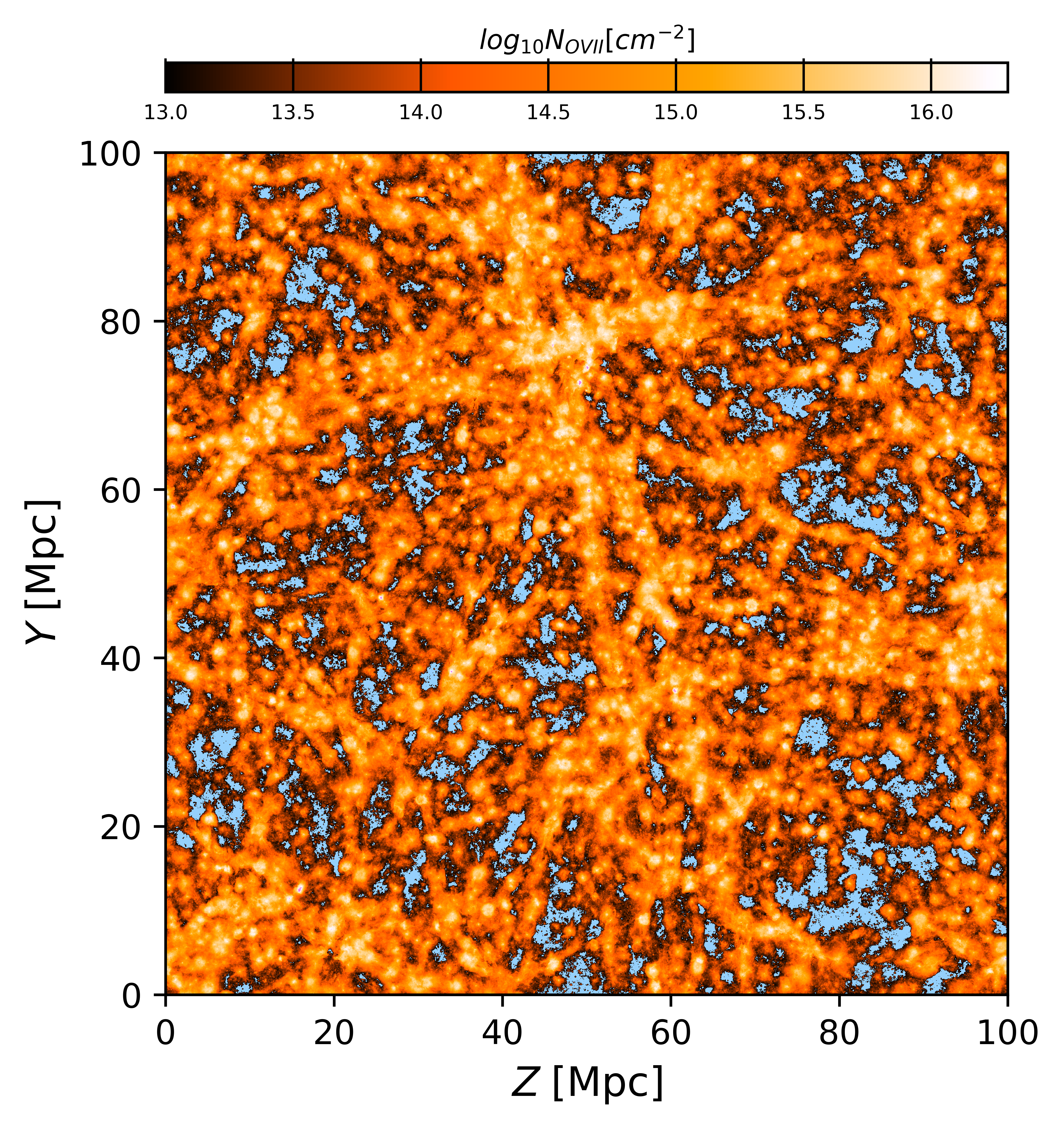}
    \end{minipage}
    \begin{minipage}{0.31\textwidth}
    \includegraphics[width=\hsize]{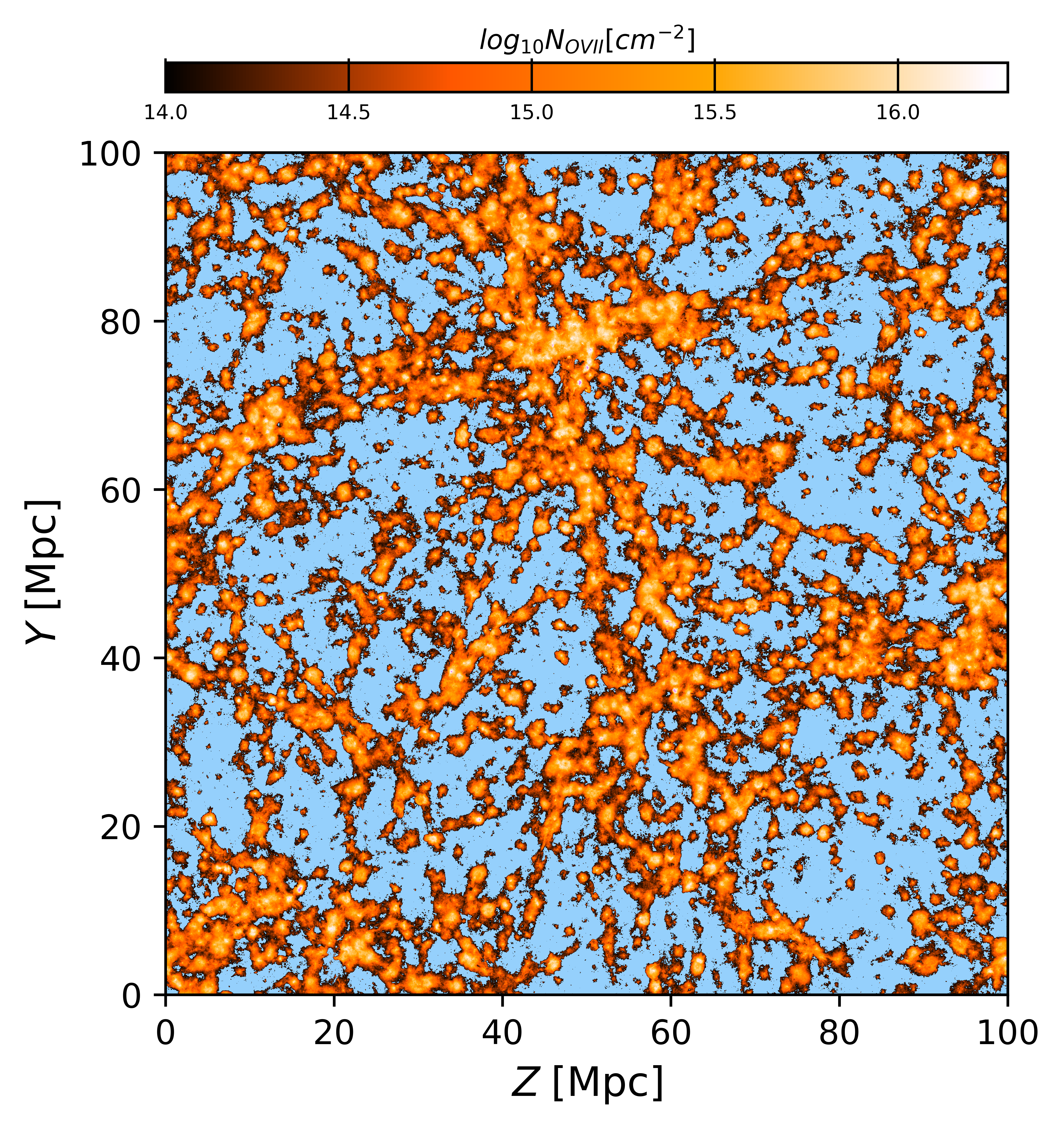}
     \end{minipage}
    \begin{minipage}{0.31\textwidth}
    \includegraphics[width=\hsize]{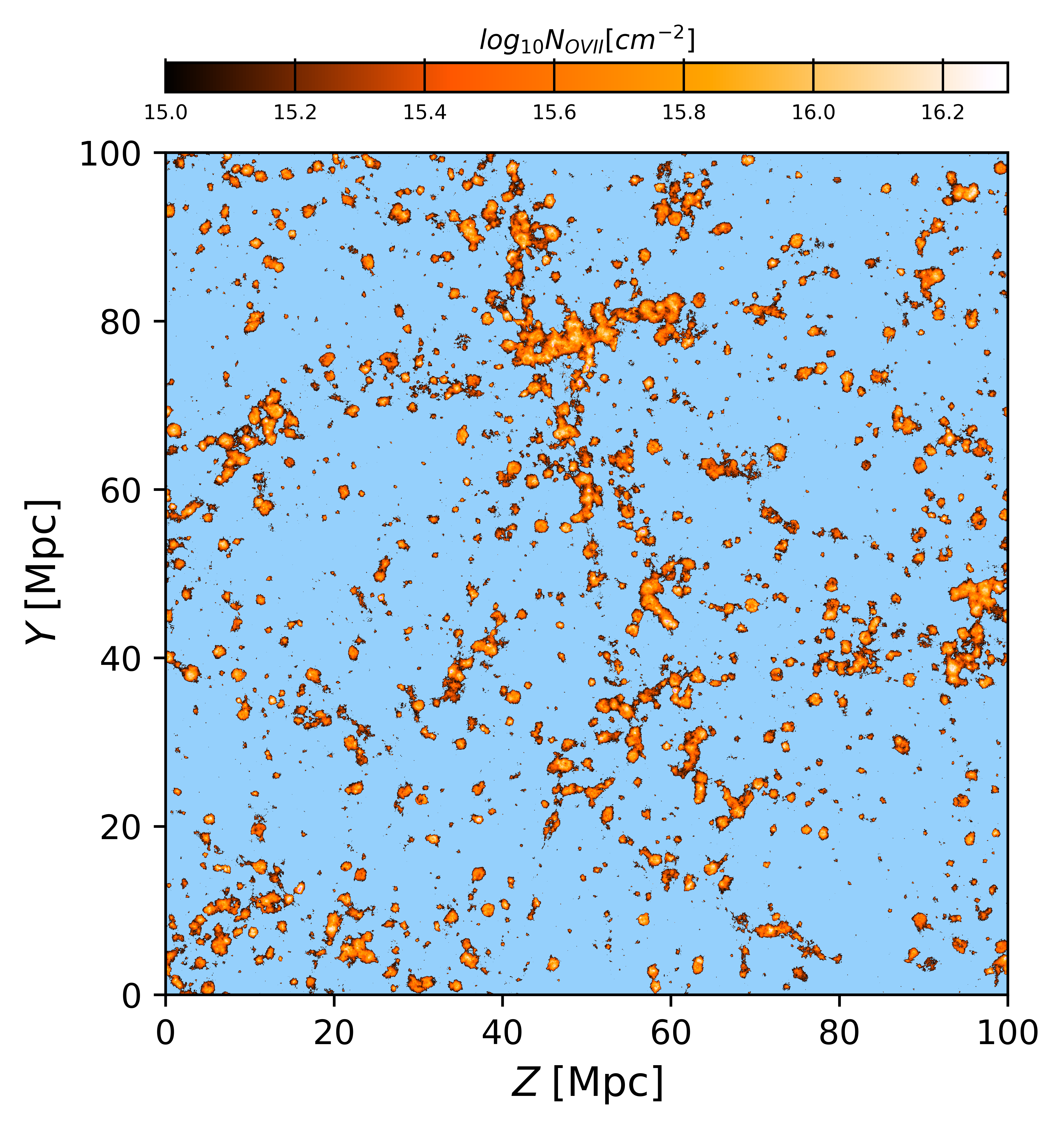}
     \end{minipage}

    \caption{Projection of the %intergalactic (i.e. outside haloes)
    $\ion{O}{VII}$ column densities across the EAGLE volume (100 Mpc), with a lower limit of $\log N(\ion{O}{VII}) > 13$ (left panel), $\log N(\ion{O}{VII}) > 14$ (middle panel), and $\log N(\ion{O}{VII}) > 15$ (right panel). 
              }
     \label{ovii_lim1314}
   \end{figure*}

\begin{figure*}
  %\vspace{-4cm}
  \hspace*{1.7cm}
%  \centering*
  \vbox{
\hbox{
\includegraphics[width=7.25cm,angle=0]{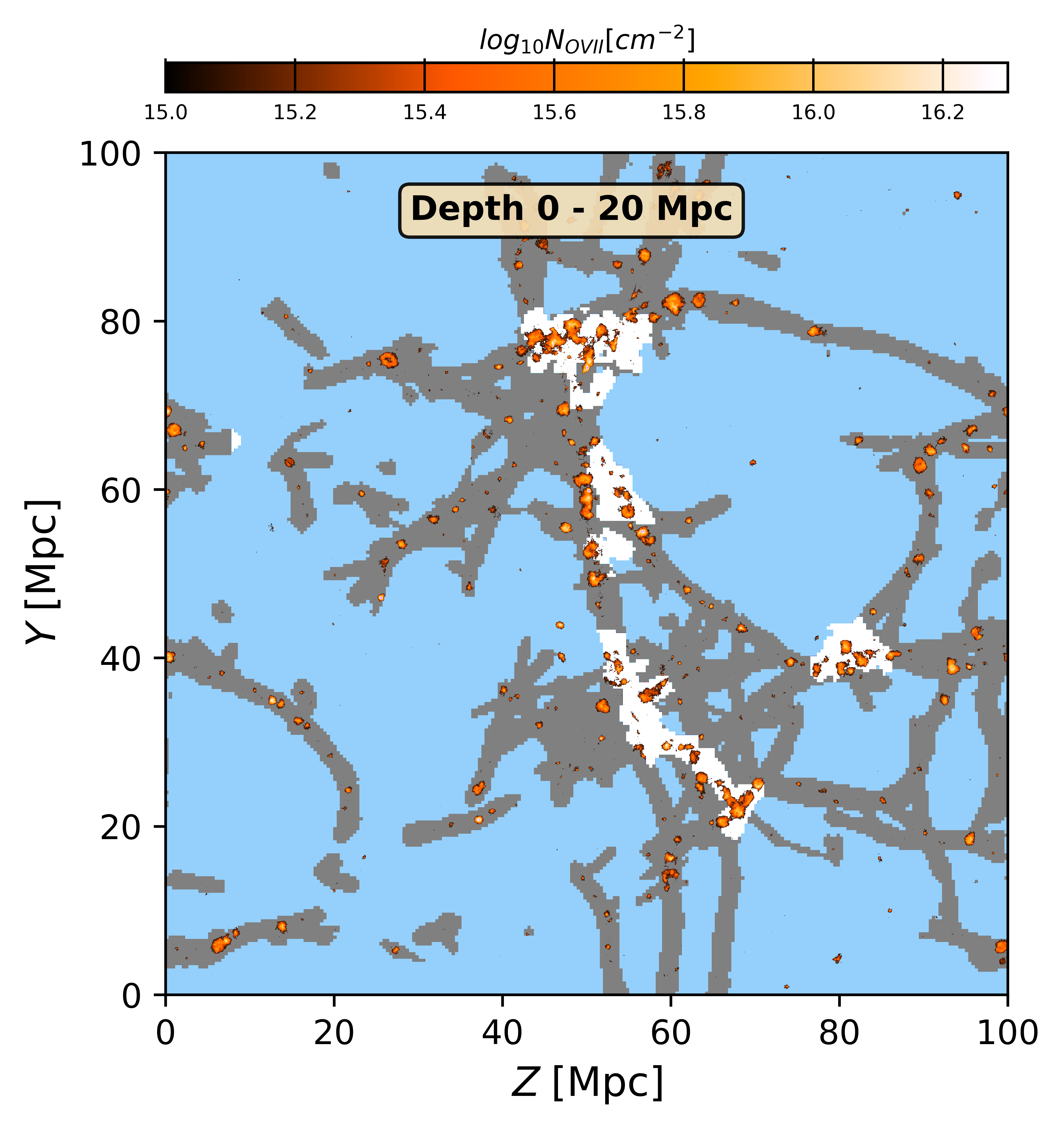}
\includegraphics[width=7.25cm,angle=0]{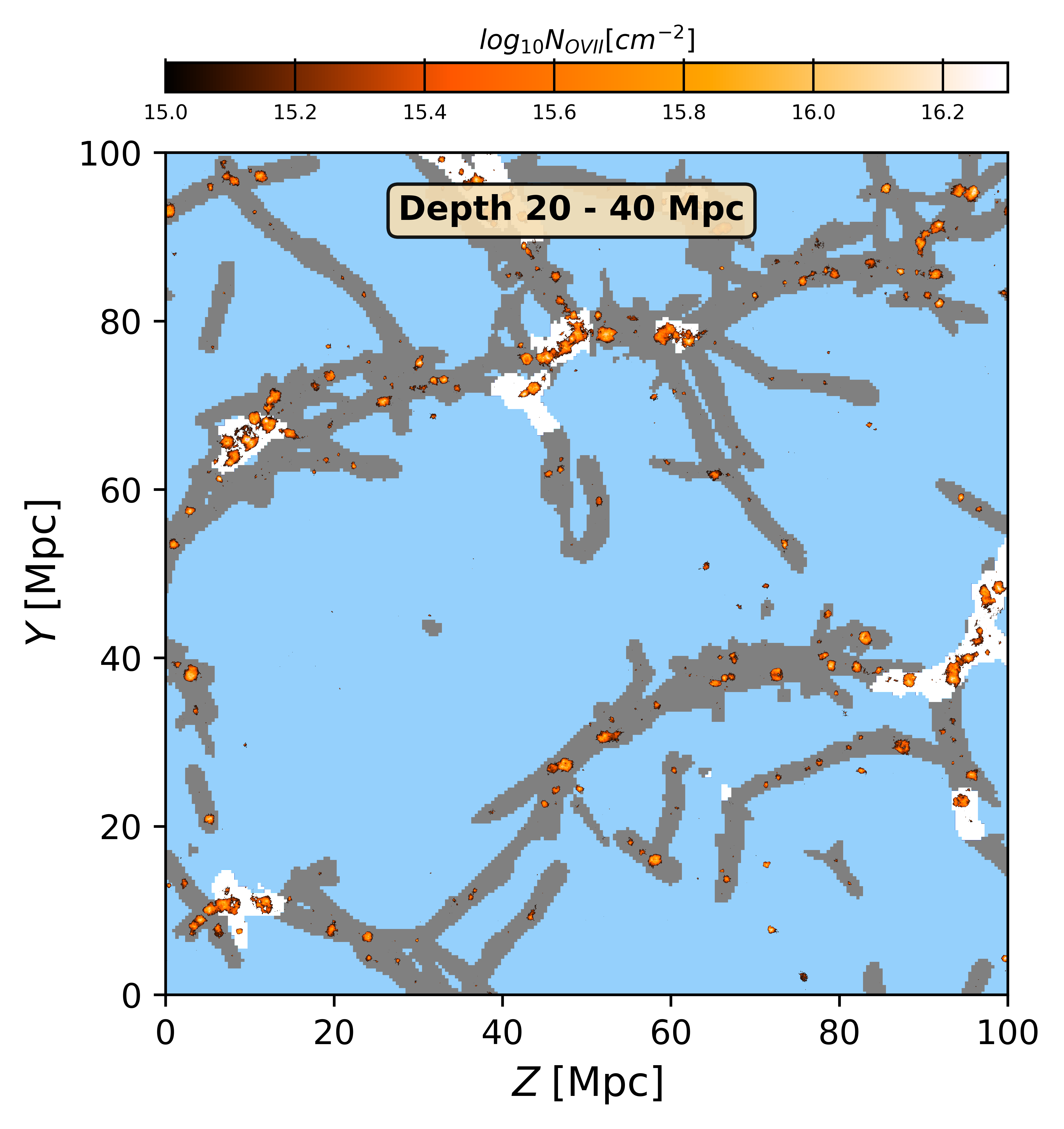}
}
\hbox{
\includegraphics[width=7.25cm,angle=0]{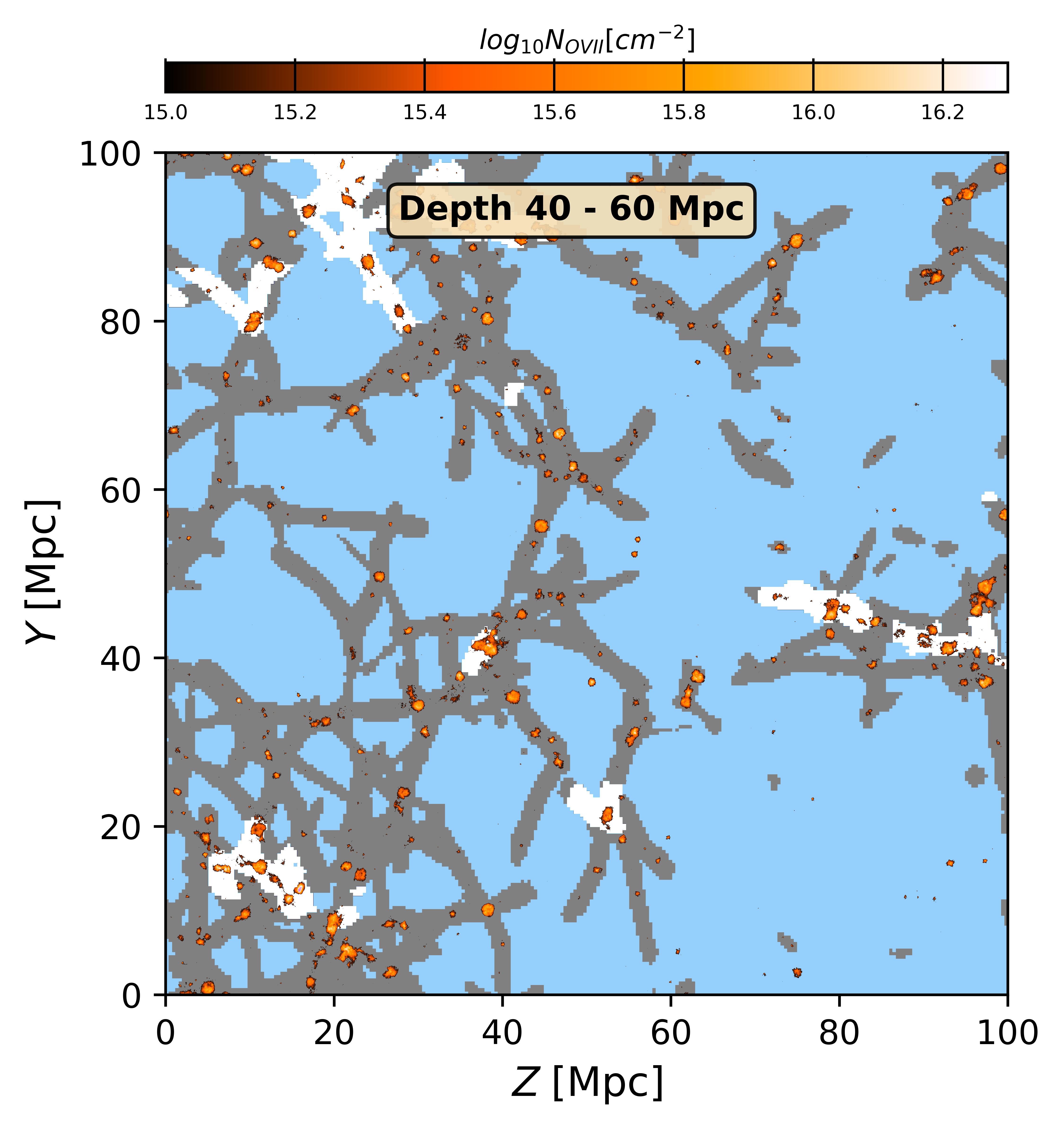}
\includegraphics[width=7.25cm,angle=0]{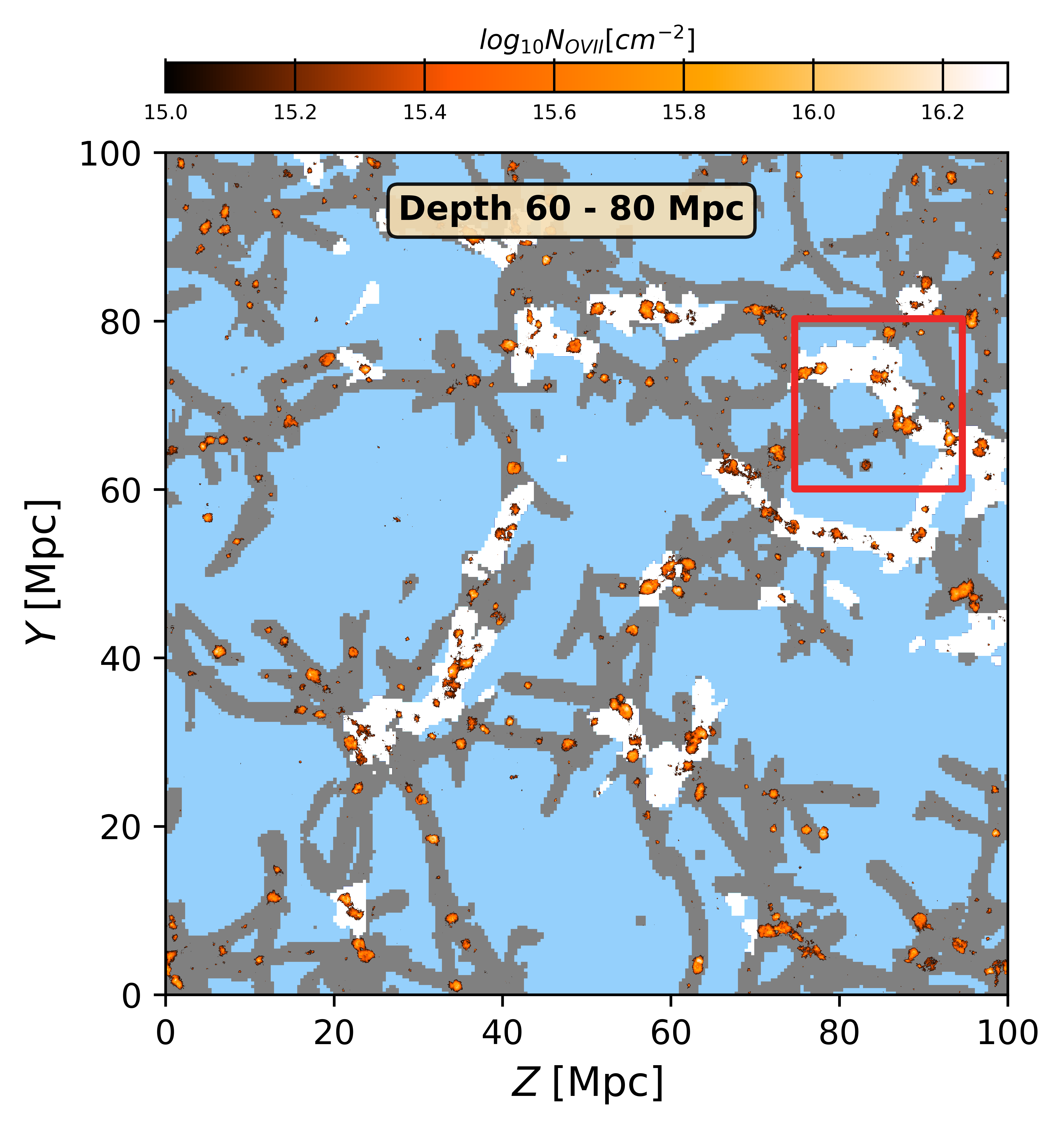}
}
\hbox{
\includegraphics[width=7.25cm,angle=0]{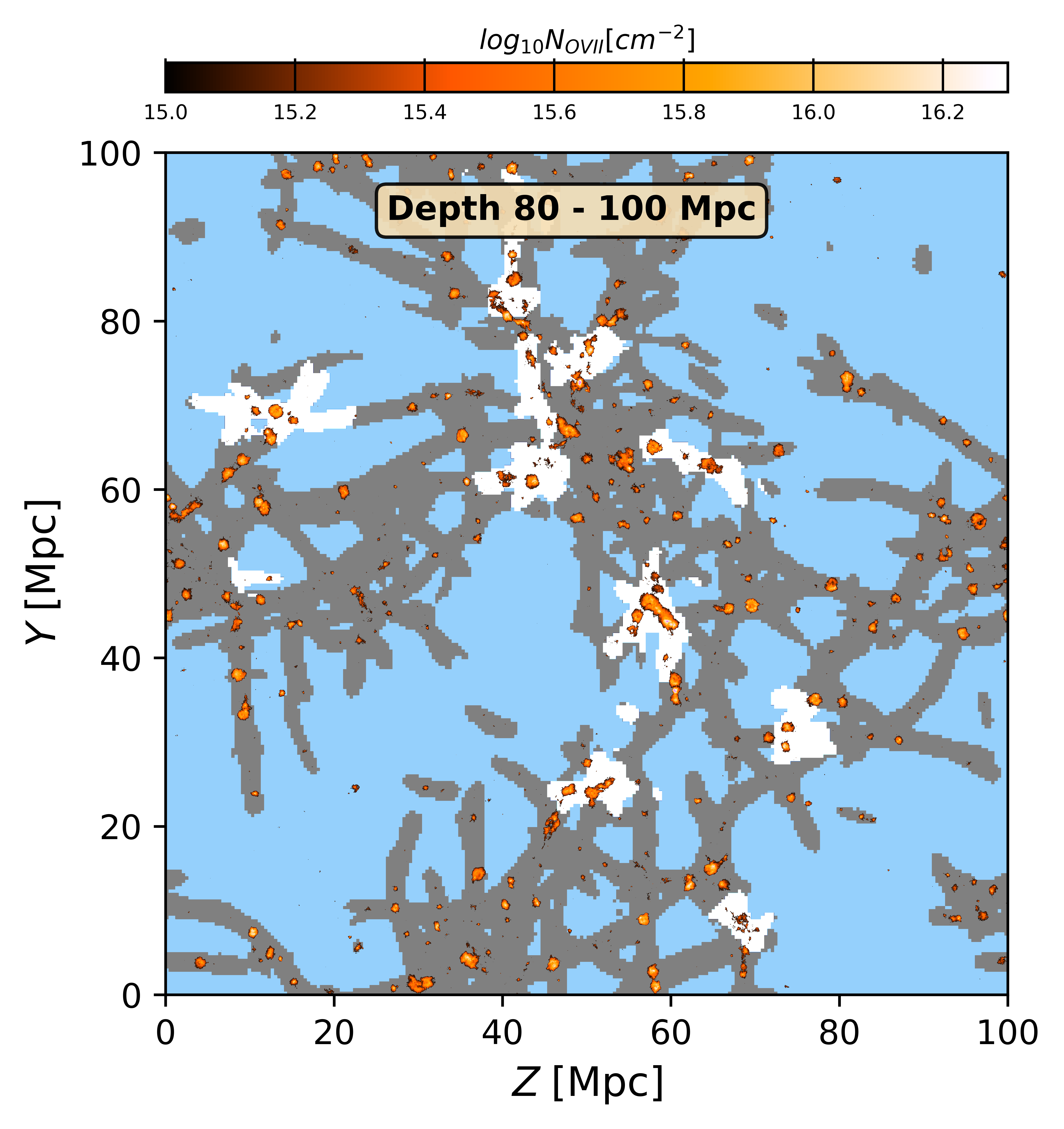}
\includegraphics[width=7.25cm,angle=0]{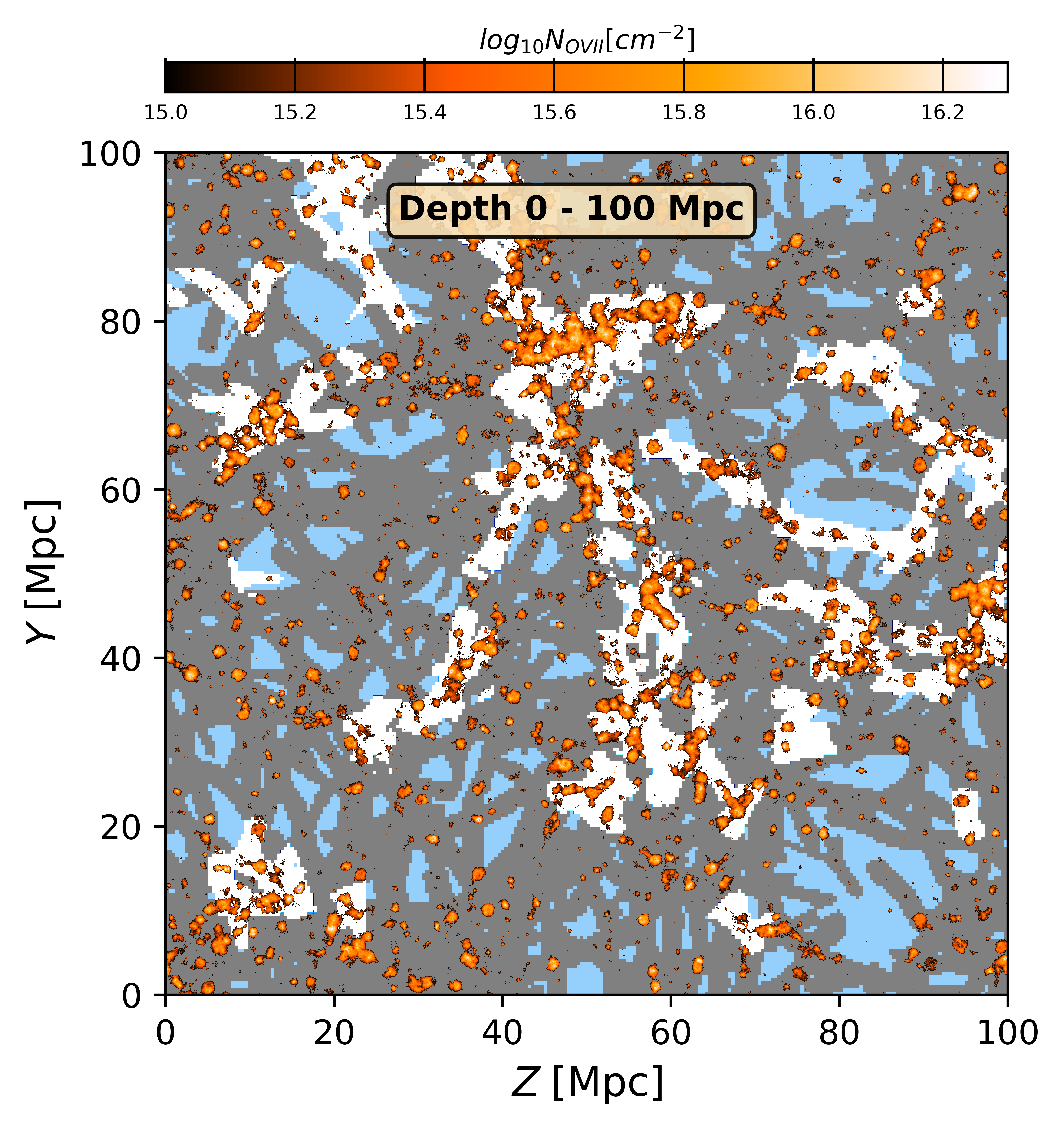}
}
}
%\vspace{-3cm}
  \caption{\ion{O}{VII} column densities (colour maps) above the approximate current and near-future X-ray detection limit of $\log N_{\ion{O}{VII}} ($cm$^{-2}) = 15$ through slices of 20 Mpc depth along the x-axis (the top and middle rows and the bottom-left panel), and the combination of all slices as a projection through the whole EAGLE box (100 Mpc, bottom-right panel). %{\bf Merkkaa kuhunkin paneliin radial range 0-20-40-60-80-100 ja 0-100} 
  %Darker and brighter shades of orange indicate lower and higher column densities, respectively. 
  The region covered by high LD Bisous filaments is indicated in white, while the grey area shows the region covered by the lower density filaments. The red square in the middle-right panel corresponds to the region in Fig. \ref{ovii_3d_plot}.}
\label{phase_slices.fig}
\end{figure*}

Ultimately, our aim was to quantify the amount of the hot WHIM traceable with \ion{O}{VII} absorption.  Indeed, most of the intergalactic \ion{O}{VII} mass is contained above the temperature limit for the hot WHIM ($\log T(K) > 5.5$) and above the approximate detection limit of $\log n_{\ion{O}{VII}} ($cm$^{3}) > -10$ (lower panel in Fig. \ref{o_ovii_phasediag}). However, since the \ion{O}{VII} density decreases rapidly as a function of distance from the nearest halo (see Fig. \ref{haloprofiles_plot}), the volume covered by \ion{O}{VII} remains low.
%(see Section \ref{ovii_fractions}). 
On the other hand, the hot WHIM density is also expected to peak around haloes. %(Tuominen+21, Fig.X).
To further investigate this, we %repeated for the 
generated a hot WHIM %the procedure to extract O and \ion{O}{VII}
density profile around haloes within a mass range of $\log M_{200}(M_{\odot}) = 12-13.5$ in the same way as for \ion{O}{VII} (see Sect. \ref{halo_profiles} for a description of the procedure). For a better comparison between hot WHIM and \ion{O}{VII}, we normalised the densities with their median value at $R_{200}$, $\log \rho_{WHIM} (M_{\odot}/Mpc^{3}) \approx 12$ for the hot WHIM and $\log n_{\ion{O}{VII}} ($cm$^{3}) \approx -9.3$ for \ion{O}{VII}. This yielded that both the hot WHIM and \ion{O}{VII} densities decrease fast with distance from the halo with a similar trend until $\approx 2 \times R_{200}$, and slightly deviate beyond that (see Fig. \ref{hW_ovii_profiles.fig}). This would imply that a large fraction of the \ion{O}{VII} and hot WHIM masses are located close to said haloes. Consequently, this suggests that we may be able to trace some of the missing baryons via \ion{O}{VII} in the surroundings of haloes. %outskirts of galaxies.

To uncover whether \ion{O}{VII} and the hot WHIM are indeed spatially connected, 
we computed the mass fraction of the hot WHIM in Bisous filaments that is contained within the same volumes as \ion{O}{VII} above $\log n_{\ion{O}{VII}} ($cm$^{3}) > -10$. 
From all the hot WHIM, $\approx 27\%$ is spatially collocated within the same volumes as \ion{O}{VII} above the aforementioned density limit. Thus, \ion{O}{VII} may be used to trace up to a quarter of the missing baryons. This, in turn, corresponds to $\approx 6\%$ of all the baryons in the EAGLE simulation. While the next generation of X-ray instruments will not be able to trace all of the missing baryons within the hot WHIM with \ion{O}{VII} absorption, they will push the limits of the observable beyond the virial radius and halo outskirts to the truly intergalactic regime. At the more realistic limit of $\log n_{\ion{O}{VII}} ($cm$^{3}) > -9$, however, the filamentary volume covered by the intergalactic \ion{O}{VII} is ten times smaller. As a consequence, the fraction of hot WHIM traced by \ion{O}{VII} is much smaller, $\approx 7\%$, which corresponds to $\approx 1\%$ of the total baryon budget. Therefore, for the hot WHIM to be properly traced by \ion{O}{VII}, filamentary haloes ought to be close enough to one another and the filament itself to be aligned with the line of sight.

 %-----------------------------------------------------------------

%-------------------------------------------------------------------------------------------------------------

%--------------------------------------------------------------------

\section{Column densities of the intergalactic \ion{O}{VII} and the detection prospects with Athena}
\label{Detection}
We utilised the above three-dimensional characterisation of the distribution of \ion{O}{VII} to obtain insights for the observational search of the intergalactic missing baryons via X-ray absorption. 
Even though the above EAGLE-based indication is that targeting filaments may only result in the tip-of-iceberg detections of \ion{O}{VII} in the galaxy outskirts, the projection of large volumes into the sight line may still make a useful X-ray observation strategy for the hot cosmic baryon gas possible.

We focus on the absorption feasibility with the X-IFU instrument on board European Space Agency's planned future X-ray mission Athena. 
Since our estimates are broad and based only on the order-of-magnitude detection limit of \ion{O}{VII}, the Athena estimates are approximately valid for the planned future missions ARCUS\footnote{http://www.arcusxray.org/}, HUBS\footnote{http://hubs.phys.tsinghua.edu.cn/en/index.html}, and Lynx\footnote{https://www.lynxobservatory.com/}, which are also expected to have similar capabilities. %\LEt{ Verify that your intended meaning has not been changed.}  
% Added by MB
Specifically, Lynx and HUBS will feature microcalorimeter arrays with spectral resolution
in the soft X--ray band that is similar to that of Athena's X-IFU \citep[$\sim 1-2$~eV resolution][]{gaskin2019,cui2020}.
Lynx will also feature a grating spectrometer with sub--eV resolution, similar to the one 
that is proposed for the Arcus mission \citep{smith2019}, both with significantly higher
resolution and collecting area than the current--generation grating spectrometers on board \textit{XMM-Newton} 
and Chandra. Thus, Arcus and Lynx may be able to extend the \ion{O}{VII} absorption work to lower column densities. We thus report the basic results also for lower detection limits in Table \ref{table:detections}.

The required 2.5 eV spectral resolution of X-IFU \citep{Nandra, Barret} corresponds to a redshift resolution of $\Delta z$ = 0.005 at 0.5 keV, which in turn corresponds to a spatial resolution of $\approx$ 20 Mpc at low redshifts. Following the procedure described in Sect. \ref{column_density}, we sampled the $\ion{O}{VII}$ column densities into 20 Mpc x 31.25$^2$ kpc$^{2}$ segments, which we interpret as individual X-IFU absorbers. 
 The result is a set of two-dimensional $\ion{O}{VII}$ column density maps (see Figs. \ref{ovii_lim1314} and \ref{phase_slices.fig}). This process was repeated along the remaining orthogonal directions (projecting along the Y- and Z-axes) to create a total of 15 column density maps.

As mentioned in Sect. \ref{ovii_fractions}, the column density maps of individual absorbers are very patchy above the limit of $\log N_{\ion{O}{VII}} ($cm$^{-2}) = 15.0$ (panels 1-5 in Fig. \ref{phase_slices.fig}). Indeed, most of the filaments (in grey and white for all and high LD filaments, respectively) are devoid of $\ion{O}{VII}$ above the detection limit. When considering a projection through the whole simulation box (100 Mpc, right panel in Fig. \ref{ovii_lim1314} and last panel of Fig. \ref{phase_slices.fig}), the picture is somewhat better, as structures from different slices are combined. But it is still far from optimal, as most of the filaments remain empty.

In order to quantify the distribution of the intergalactic $\ion{O}{VII}$ absorbers 
(see above for the definition of an absorption system used here), we computed its column density distribution function (CDDF). The CDDF is defined as follows:

\begin{equation}
    f(N,z) \equiv \frac{\partial^{2} n}{\partial \log N \partial z},
\label{CDDF.eq}
\end{equation}
\\
where $n$ corresponds to the number of absorbers, $N$ is the column density (in our case of $\ion{O}{VII}$) and $z$ is the redshift. 
To investigate the environmental distribution of the $\ion{O}{VII}$ absorbers in the Cosmic Web, we calculated the CDDFs for (1) all the $\ion{O}{VII}$ absorbers, (2) intergalactic $\ion{O}{VII}$ within Bisous filaments, and (3) $\ion{O}{VII}$ within FoF haloes in Bisous filaments. 

\begin{figure}
   \centering
   \includegraphics[width=\hsize]{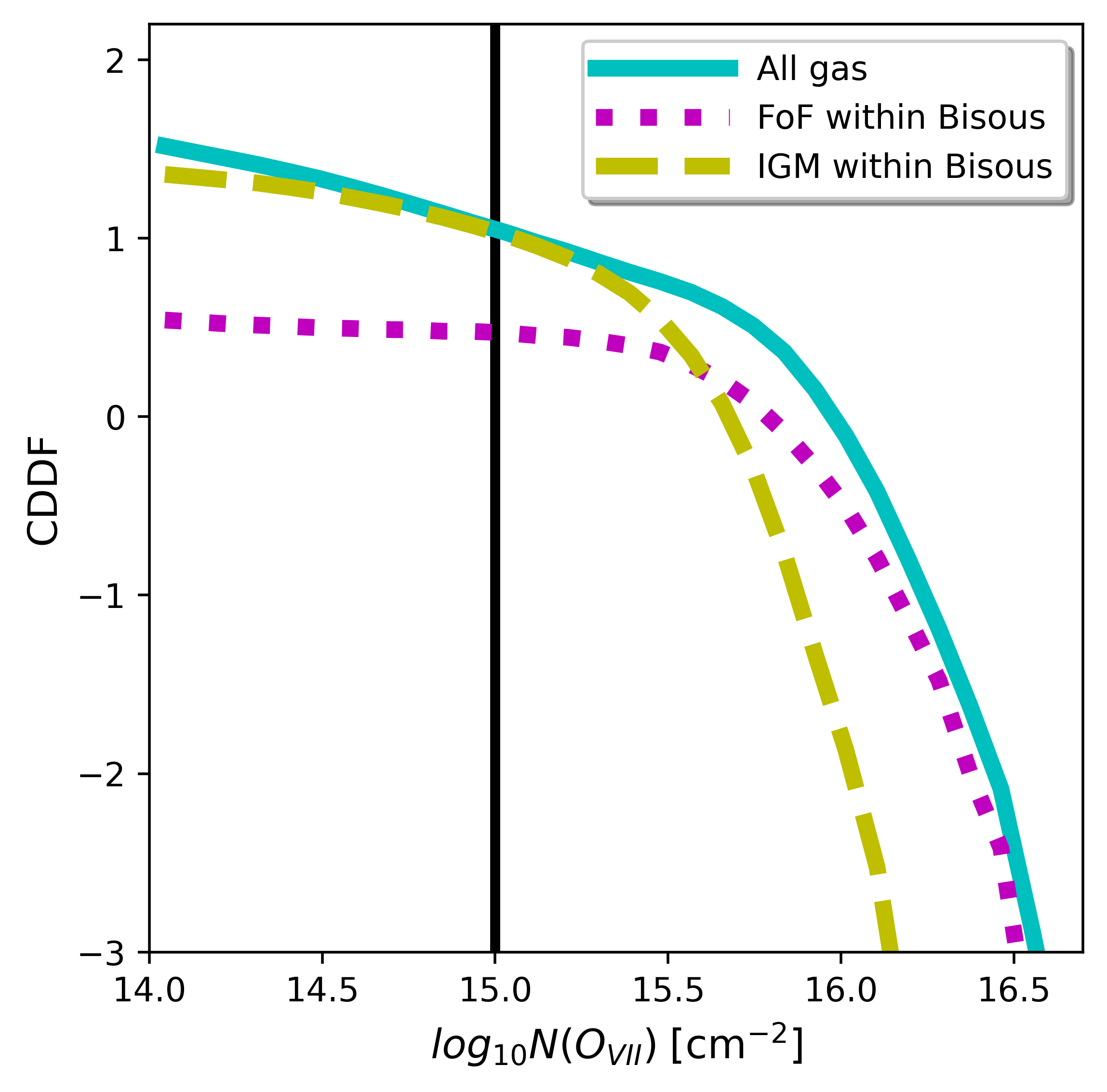}
      \caption{ CDDF of EAGLE  $\ion{O}{VII}$ absorbers (1) in the full simulation (cyan line), (2) within FoF haloes in Bisous filaments (dotted purple line), and (3) within the IGM in Bisous filaments (dashed yellow line). The vertical black line indicates the approximate detection limit of current and near-future X-ray instruments. %{\bf tee cumulative}
              }
         \label{cddf}
   \end{figure}

We found that at the low values the distributions decrease relatively slowly with increasing column density but beyond a break at $\log N_{\ion{O}{VII}} ($cm$^{-2}) \approx 15.8$ they decrease  steeply %exponentially 
(see Fig. \ref{cddf}).
The $\ion{O}{VII}$ column densities below the break are dominated by the intergalactic gas while above the break the dominating source is the gas within haloes. Indeed, \citet{2020MNRAS.498..574W} describe the break (or `knee') 
as a transition from gas within haloes to the IGM. While they concentrated on the halo component, our focus here is the IGM within filaments.

In order to derive properly the cosmologically interesting information about the number of absorption systems per unit redshift $dz$ one should analyse a deep enough light cone simulation so that (1) the sampled volume is substantial in order to minimise the cosmic variance and (2) the possible redshift evolution of the absorbers would be included. The (100 Mpc)$^3$ snapshot at z=0 studied in this work has very limited use for this purpose. However, for comparison with other works, keeping the above caveats in mind, we performed the exercise of integrating the CDDF (Eq. \ref{CDDF.eq} and Fig. \ref{cddf}). We found that when considering all intergalactic \ion{O}{VII}, there are $\approx$ 4 absorbers per unit redshift at $\log N_{\ion{O}{VII}} ($cm$^{-2}) > 15$.  The comparison of this value to other works is complicated due to, for example, varying definitions of an absorber and treatments of the haloes and intergalactic matter. Yet, our estimate is similar to those obtained by \citet{2009ApJ...697..328B}, \citet{2006ApJ...650..560C}, and \citet{2019MNRAS.488.2947W}.
%\LEt{ When listing references in the running text of a sentence, please separate by commas and place an "and" between the last comma and the reference (or an "and" and no comma if there are only two references). Please check for this throughout the paper.} 

As discussed in Sect. \ref{ovii_fractions}, most of the intergalactic \ion{O}{VII} lies within filaments. What is more, around the observable level of $\log N_{\ion{O}{VII}} ($cm$^{-2}) > 15$ the intergalactic component of the CDDF is completely dominated by gas within Bisous filaments. This means that virtually all observable intergalactic $\ion{O}{VII}$ is located within filaments.

In the following sections we provide observational predictions for these filamentary $\ion{O}{VII}$ column densities in two different large-scale surveys, SDSS and 4-metre Multi-Object Spectroscopic Telescope (4MOST) survey.

\subsection{SDSS}

Given that the observational X-ray absorption signals are presumably very weak, it is important to detect the filaments harbouring the hot WHIM in optical surveys. This provides the redshift for the absorber, which helps identify the spectral line and to constrain the physical properties of the absorber.   
In addition, the knowledge of the large-scale environment aids in confirming the intergalactic nature of the absorber.

We consider first 
%currently largest well-controlled
the Bisous filament catalogue of \citet{2014MNRAS.438.3465T}, constructed from SDSS galaxies, which is quite complete up to redshift $z$=0.05, corresponding to a depth of $\approx$ 215 Mpc. %(see Section \ref{filaments}). 
The filament catalogue starts from a radial distance of $\approx 85$ Mpc in order to avoid the Local Void, an under-dense region surrounding our Milky Way. %{\bf (Tempel...)}. 
Thus, the radial depth of the filament sample is $\approx 130$ Mpc, slightly larger than the EAGLE simulation box.

In order to quantify the feasibility of the observational search for the missing baryons
we first calculated the fraction of the SDSS footprint covered by filaments.  
To this end, we randomly selected 10000 sight lines through the plane of sky covered by SDSS and found that a vast majority ($\approx$ 92\%)  of such sight lines cross a filament at least once (see Fig. \ref{SDSS_bis}, bottom-left panel). When considering only the high LD filaments (i.e. filaments harbouring a larger fraction of missing baryons; see Sect. \ref{filaments}), this covering area is $\approx$23\% of the whole survey (Fig. \ref{SDSS_bis}, bottom-right panel).

In order to estimate how accurately the observational SDSS filament network geometry (i.e. number densities and extents of the filaments) are reproduced by the EAGLE simulations, we repeated the area covering analysis for filaments in the simulation. We applied the same Bisous method for detecting both the observational and the simulated filaments. %(see Section \ref{filaments}).
To this end, we again selected 10000 random sight lines crossing the simulation through a random orthogonal direction (X, Y or Z), and adjusted for the difference in the SDSS and EAGLE depths.
   Of all the sight lines, $\approx$ 91\% and $\approx$ 23\% cross at least once a filament or a high LD filament, respectively (Fig. \ref{SDSS_bis}, top panels).% and Table X).
Thus, the projected areas covered by filaments in SDSS and EAGLE are consistent within $\sim$1\%.  The excellent agreement gives us confidence that 
the basic geometric properties of filaments (the number density and the extent) are accurately reproduced in the EAGLE simulations. This justifies the usage of EAGLE filaments as a basis for making predictions for the baryon content of the Cosmic Web filaments.

\begin{figure*}
\begin{minipage}{0.48\textwidth}
\centering
\includegraphics[width=\hsize]{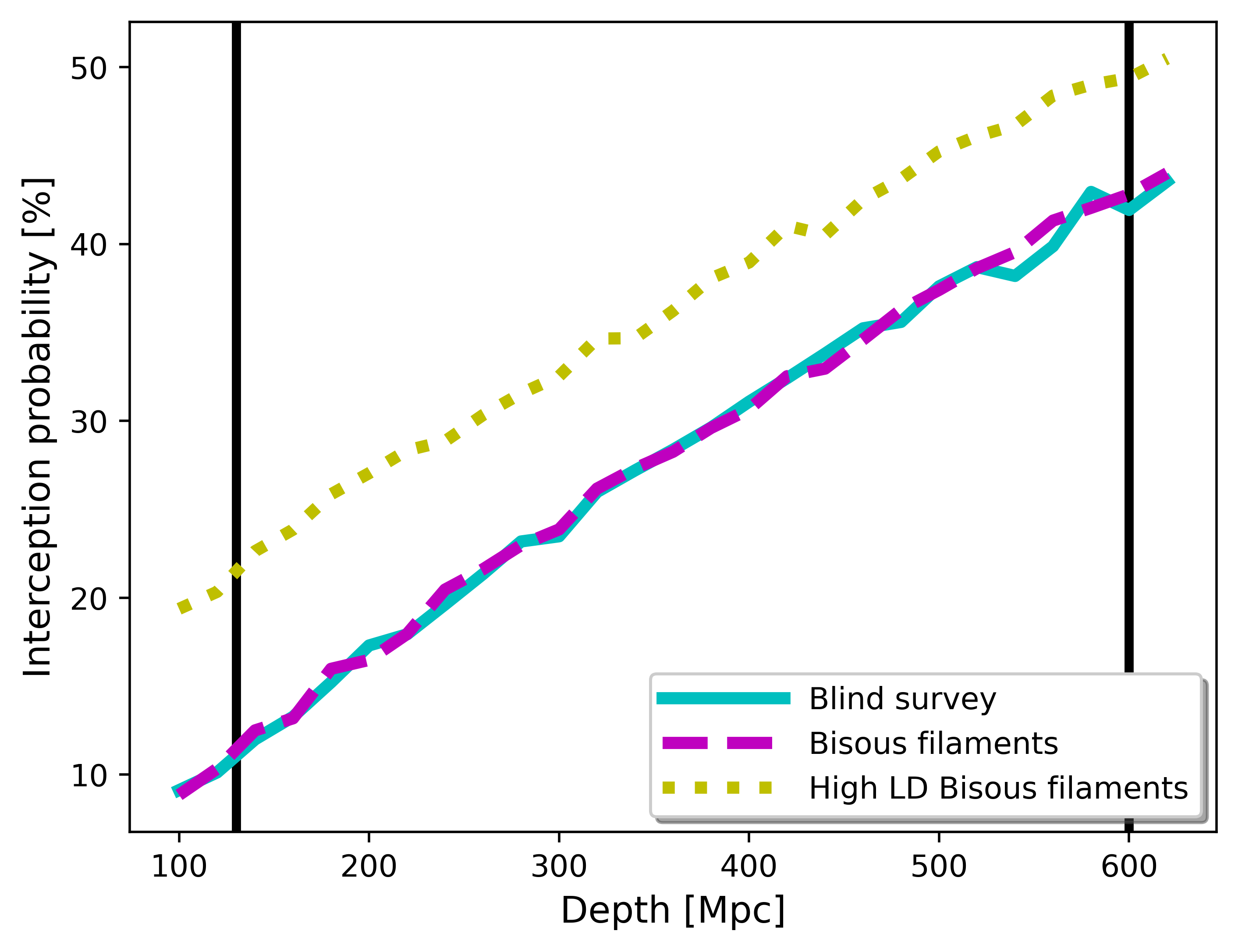}
\end{minipage}
\begin{minipage}{0.48\textwidth}
\centering       
\includegraphics[width=\hsize]{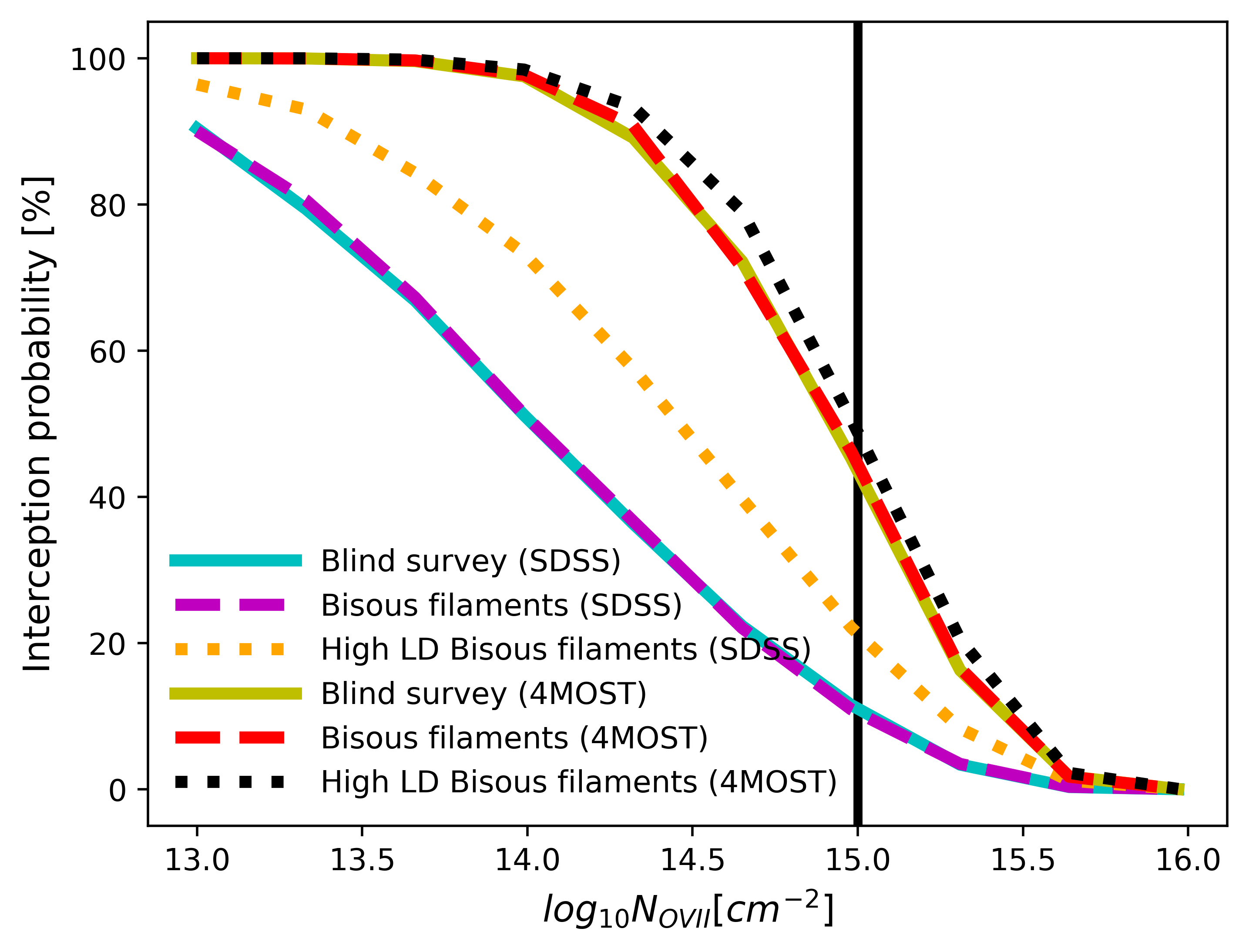}
\end{minipage}
         \caption{EAGLE-based probability of intercepting an intergalactic \ion{O}{VII} absorber (see Sect. 5 for our definition of such an absorber). {\it  Left panel:}  Probability for an absorber with column density $\log N(\ion{O}{VII} ($cm$^{-2}) > 15$ indicated as a function of the probed path length (1) for a blind search (solid turquoise line), (2) when targeting Bisous filaments (dashed purple line), and (3) when targeting high LD Bisous filaments (dotted orange line). The vertical black lines indicate the depths of the SDSS survey (left) and 4MOST 4HS survey (right). %{\bf  Kirjoita "SDSS" ja "4MOST" kuvaan mustan viivan viereen}
         {\it  Right panel:} Interception probability in different filament targeting scenarios along a 130 (600) Mpc path corresponding to the SDSS (4MOST) samples discussed in the text: no targeting (solid turquoise and yellow lines%\LEt{ We reserve the use of slashes to denote ratios and instrument pairings and for use in equations.}
         ), targeting Bisous filaments (dashed  purple and red lines), and targeting high LD Bisous filaments (dotted orange and black lines). The vertical solid black line indicates the approximate detection level of current and near-future X-ray absorption instruments.}
\label{interception_percentage}
\end{figure*}

\begin{figure*}
    \vbox{
    \hbox{
    \begin{minipage}{0.48\textwidth}
    \centering   
    \includegraphics[width=\hsize]{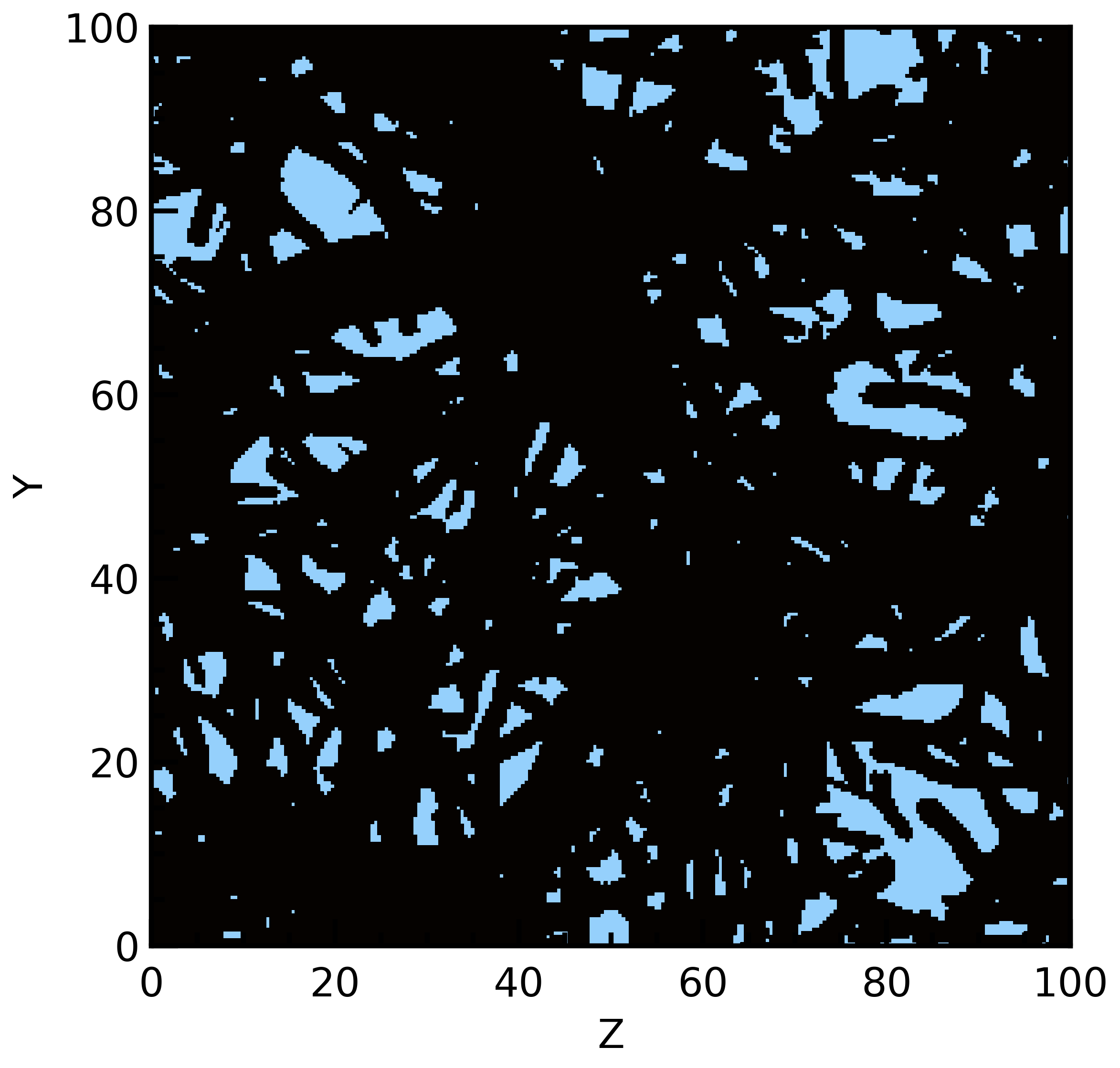}
           
    \end{minipage}
    \begin{minipage}{0.48\textwidth}
    \includegraphics[width=\hsize]{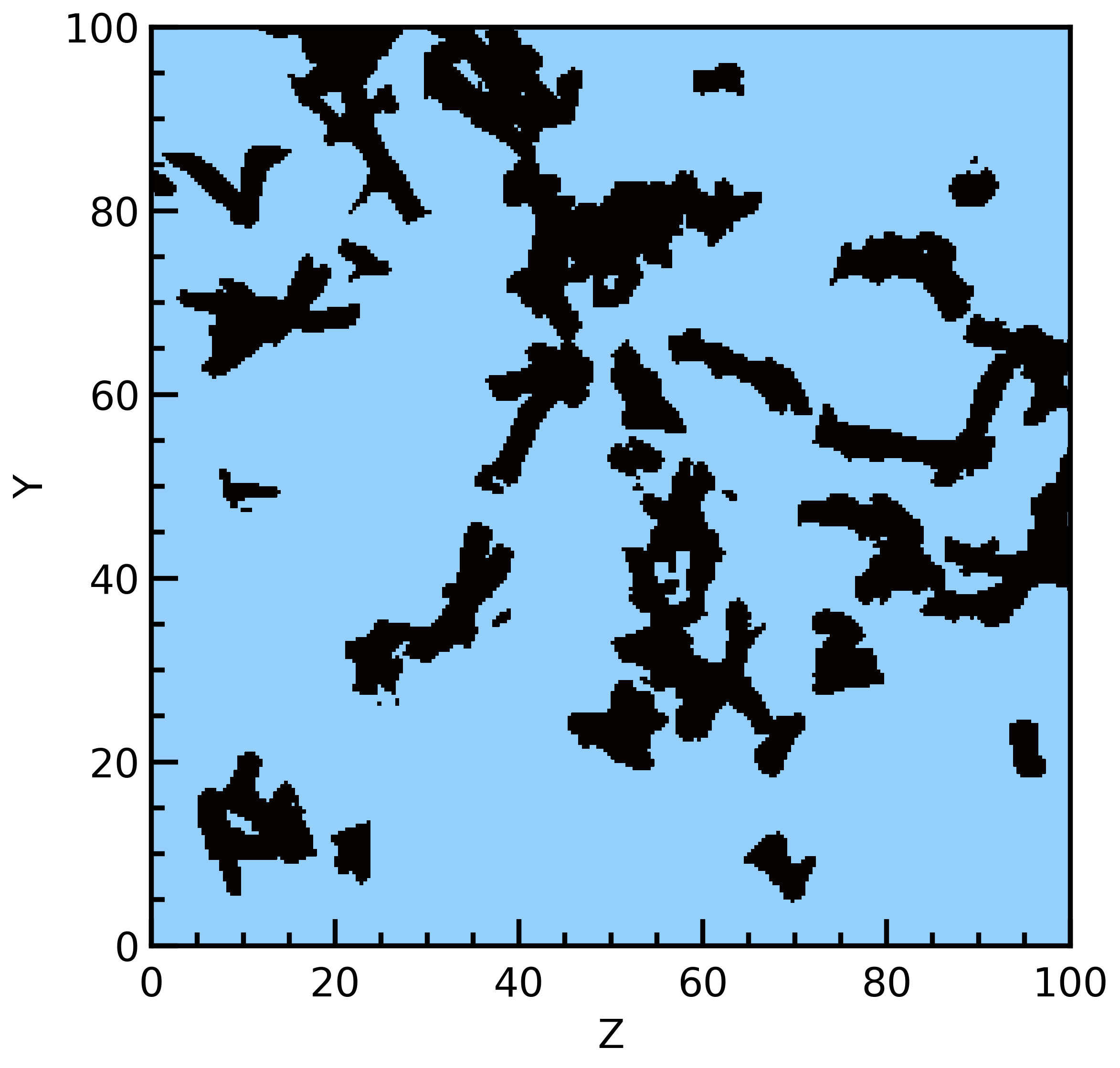}
    \end{minipage}
    }
    \hbox{
    \begin{minipage}{0.48\textwidth}
    \centering   
    \includegraphics[width=\hsize]{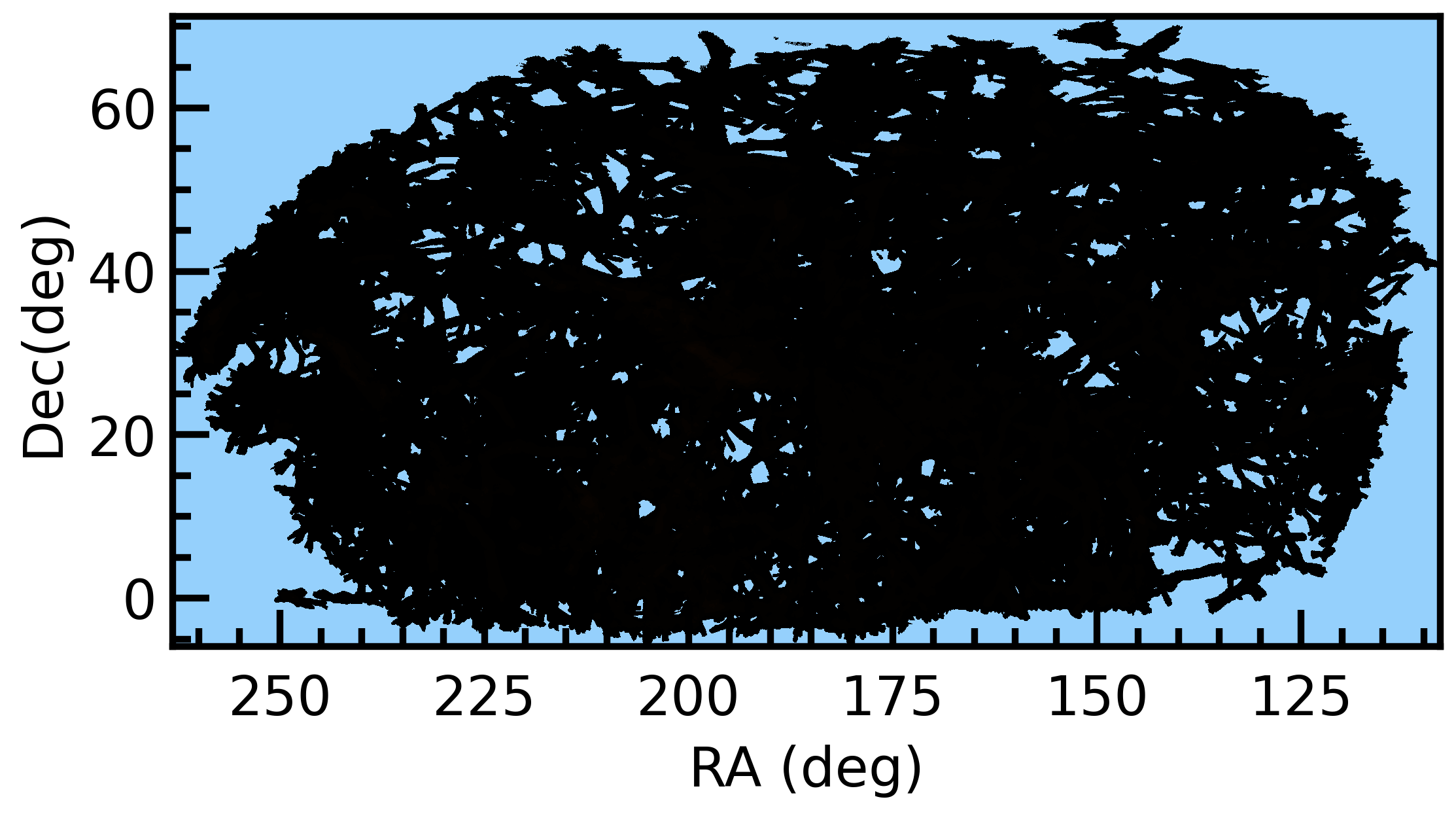}
         
    \end{minipage}
    \begin{minipage}{0.48\textwidth}
    \includegraphics[width=\hsize]{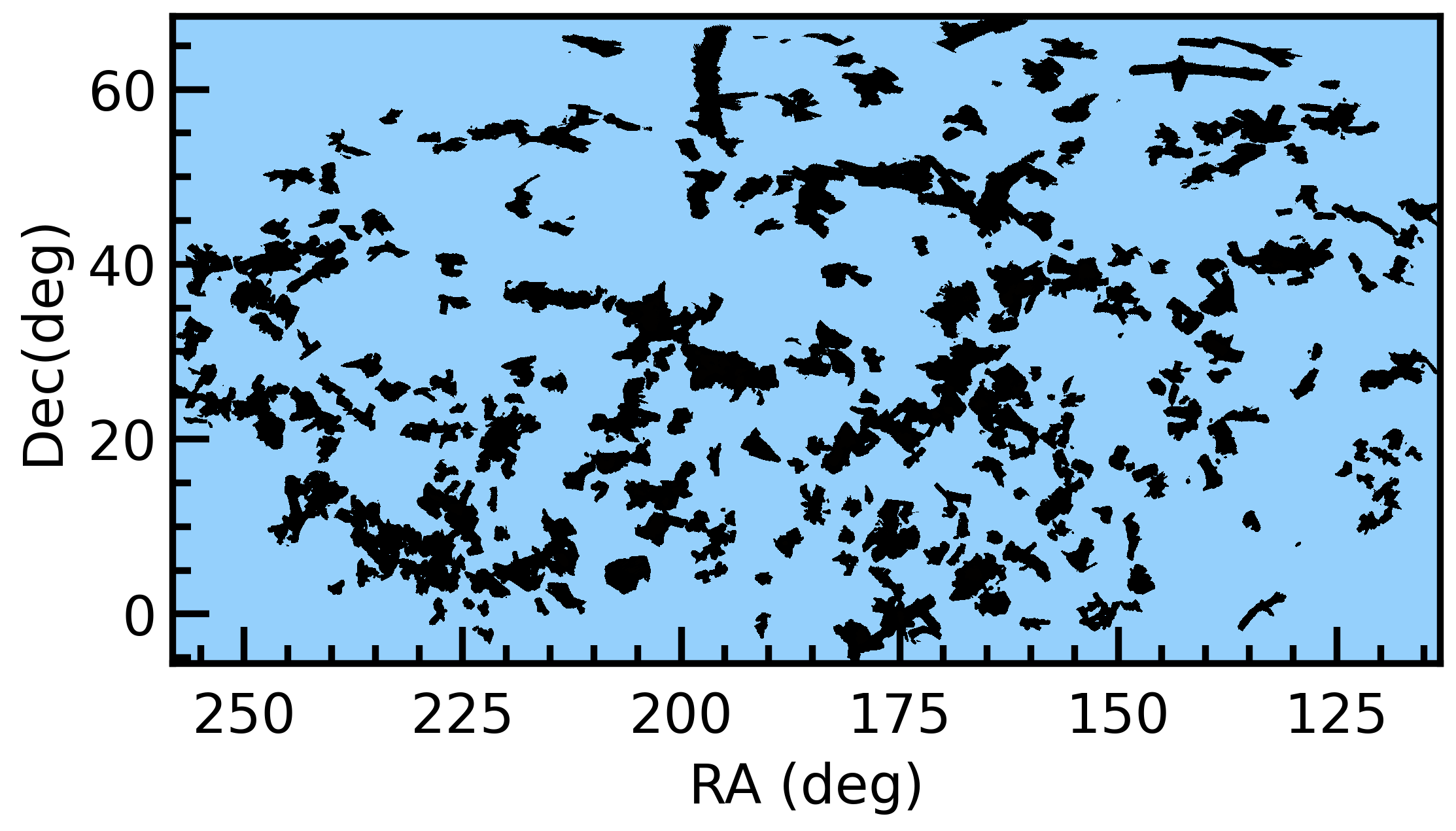}
    \end{minipage}
    }
    }
    \caption{Area covered by Bisous filaments in EAGLE and SDSS. \textit{%\LEt{ Please provide a title for this figure.}
    Upper-left panel:} Projection of the Bisous filaments (black) along the X-direction through the whole EAGLE simulation box (100 Mpc). \textit{Upper-right panel:} Same as on the left, but selecting only filaments within high LD regions. \textit{Lower-left panel:} Bisous filaments in the SDSS main footprint (black) projected on the plane of sky in the radial range $z$ = 0.02-0.05, or $\approx$ 85-215 Mpc from \citet{2014MNRAS.438.3465T}. \textit{Lower-right panel:} Same as on the left, but selecting only filaments within high LD regions.
              }
     \label{SDSS_bis}
   \end{figure*}

 We then proceeded to examine the column densities along the random sight lines to find out the number of instances where the values exceed $\log N_{\ion{O}{VII}} ($cm$^{-2}) = 15$, the approximate sensitivity limit of X-IFU set by the systematics related to the spectral resolution \citep{2013arXiv1306.2324K}. As described above,  the simulation box was divided into 15 segments (5 in %\LEt{ in?}
 each orthogonal direction) 
 of 20 Mpc length, corresponding to the X-IFU resolution. To derive predictions for SDSS we selected 6 segments for a depth of 120 Mpc, an approximation for the survey depth \newline (130 Mpc). In practice, each of the 10000 sight lines first crossed the whole simulation box in one random orthogonal direction and then through one segment towards a different direction.
We found that in the corresponding sample $\approx$ 11\% of the sight lines contain at least one absorber with $\log N_{\ion{O}{VII}}$(cm$^{-2}$) $>$ 15. We interpret this value as the probability $P_{N(\ion{O}{VII})}$ of encountering at least one intergalactic absorber with $\log N_{\ion{O}{VII}}$(cm$^{-2}$) $>$ 15 per random sight line through SDSS with X-IFU's resolution (see Fig. \ref{interception_percentage}, left panel).
%{\bf The outlook is quite pessimistic... the probability maintains low even if focusing on the high LD filaments...}

The most optimal background X-ray objects for producing the spectra to be absorbed by the intervening WHIM are presumably blazars.  
The current Athena Mock Observation Plan contains 39 AGN dedicated to the WHIM absorption study. The exposures are planned so that the statistical sensitivity of the X-ray signal at 20 \AA\ is at the level of the expected systematic uncertainties (i.e. $\log N_{\ion{O}{VII}}$(cm$^{-2}$) $\sim$ 15).
The AGN list is not selected based on the foreground structure and thus may be considered as a `blind' search. The above $P_{N(\ion{O}{VII})}$ $\approx$ 11\% thus applies to the sight lines towards the $\sim$10 
%{\bf (Toni, tsekkaa moniko noista on SDSS:n alueella)} 
AGN covered by SDSS. Thus, we expect $\sim 1$ intergalactic \ion{O}{VII} detection with Athena X-IFU in the SDSS region with the currently planned AGN.

To improve the detection probability, we investigated the feasibility of targeting the optically detected filaments with Athena 
as an alternative to the blind search.
However, as described above, the sky coverage of the full filament sample is above 90\% (see the upper-left panel of Fig. \ref{SDSS_bis}), and thus the probability, $P_{N(\ion{O}{VII})}$, of detecting at least one absorber along the line of sight does not improve when focusing on filaments. 
We thus investigated the strategy of targeting the filaments with the highest expected WHIM density, as indicated by the highest galaxy LD \citep{2021A&A...646A.156T}. 
We repeated the above calculations, this time selecting only such sight lines, which cross any high LD filament (see Fig. \ref{SDSS_bis}).
In this strategy the probability of encountering at least one absorber above Athena detection limit per sight line increases from the blind search value of $\approx$ 11\% to $\approx$ 22\% (see Fig. \ref{interception_percentage}).  
While the probability doubles, the improvement remains modest due to relatively poor correlation between LD and \ion{O}{VII} density (see Fig. \ref{phase_slices.fig}).

\subsection{4MOST}

The 4-metre Multi-Object Spectroscopic Telescope \citep[4MOST;][]{2019Msngr.175....3D} is a wide-field spectroscopic facility aimed at providing large and deep surveys of the southern sky (under development at the time of writing) \footnote{https://www.4most.eu}. One community survey in particular, the 4MOST Hemisphere Survey of the Nearby Universe (4HS), will extend the depth of a complete galaxy survey up to $z = 0.15$ (i.e. $\sim 600$ Mpc) with a similar quality to the SDSS sample at $z = 0.05$ (described in the previous section). The higher depth of the survey will allow the construction of a robust filament sample up to $z = 0.15$. This increases the probability of randomly crossing at least one absorber of $\log N_{\ion{O}{VII}} ($cm$^{-2}) > 15$ per sight line from $\approx$11\% in SDSS to $\approx$45\% (see Fig. \ref{interception_percentage}). 
Targeting the high LD filaments slightly increases the probability of
crossing one absorber of $\log N_{\ion{O}{VII}} ($cm$^{-2}) > 15$ per sight line to a comfortable level of $\approx$50\%.

The important improvement afforded with the 4MOST 4HS survey is the expected large ($\approx$50\%) sky coverage up to a redshift depth of $z = 0.15$.
Combining this with our EAGLE-based estimate that $\approx$ 80\% of the 4HS depth will be covered by high LD filaments, 
we expect to cover 0.5 $\times$ 0.8  = 40\% of the full sky with high LD 4HS filaments. This would enable the selection of a substantial number of the brightest blazars behind these structures. We will study this issue further with the blazar flux statistics in a future work.

%-------------------------------------------------------------
%
\begin{table}
\caption{Probability, $P_{N(\ion{O}{VII})}$,  of intercepting at least one absorber above a given column density limit per sight line for a depth of 120 and 600 Mpc, which corresponds to the SDSS and 4MOST 4HS surveys, respectively. %targeting all filaments and
The probability of intercepting one absorber targeting high LD filaments is indicated in brackets. 
%{\bf SDSS depth on 130 Mpc, ei 200. Miksi kaksi eri arvoa X/Y?}    
    }         % title of Table
\label{table:detections}      % is used to refer this table in the text
\centering                          % used for centering table
\begin{tabular}{c c c}        % centered columns (4 columns)
\hline\hline                 % inserts double horizontal lines
   min. $\log N_{\ion{O}{VII}}$  & 120 Mpc & 600 Mpc \\    % table heading 
\hline                        % inserts single horizontal line
14.0   & 50\% (72\%) & 98\% (98\%) \\      % inserting body of the table
14.5  & 28\% (48\%) & 80\% (85\%)    \\
15.0 &  11\% (22\%) & 45\% (50\%)     \\
15.5   & 1.5\% (4\%) & 8\% (10\%) \\
\hline                                   %inserts single line
\end{tabular}
\end{table}
%
%-------------------------------------------------------------
%                                             Two column Table 

\subsection{Limitations of simulations}  
As described above, the probability, $P_{N(\ion{O}{VII})}$, of intercepting at least one $\ion{O}{VII}$ absorber per sight line is rather low for absorbers above the column density limit of $\log N(\ion{O}{VII}) > 15$, when considering the area covered by SDSS and the depth through which the Bisous filament sample is complete ($\approx 130 $Mpc). However, this probability is highly dependant on the column density (right panel in Fig. \ref{interception_percentage}).
 At slightly lower column densities, $\log N(\ion{O}{VII}) \ge 14.5$, the interception probability increases to $\approx$ 45\%  ($\approx$ 60\%) from $\approx 11\%$ ($\approx 22\%$) at $\log N(\ion{O}{VII}) > 15$ when targeting all (high LD) SDSS filaments, respectively 
 (see Table \ref{table:detections} and Fig. \ref{ovii_lim1314} for a projection of $\ion{O}{VII}$ at lower column densities).
The deeper surveys with 4MOST would increase the interception probability further to $\approx 80\%$ ($\approx 85\%$)
at $\log N(\ion{O}{VII}) \ge 14.5$ for the full (high LD) filament samples, respectively. These levels would perhaps enable an efficient X-ray search for %\LEt{ for?}
the absorbers. 

Given that the detection probabilities are very sensitive to the simulation results around $\log N_{\ion{O}{VII}}(cm^{-2}) \sim 15$, it is important to understand how reliable the EAGLE results are. 
As discussed by \citet{2022MNRAS.511.2600M}, the metal transportation is very model dependant, since different simulations implement different sub-grid processes for the stellar and AGN feedback. Thus, it would be useful to compare the results of our work with those obtained using different simulations. However, different published works have different scientific aims and scopes and consequently detailed one-to-one comparison of the spread of intergalactic \ion{O}{VII} is currently not feasible. In the following we place our results in the context of other simulations to the extent allowed by the current literature.

\citet{2022MNRAS.511.2600M} studied the transportation of metals  within and beyond haloes in EAGLE. They found that the mass of metals ejected outside R$_{200}$ peaks at halo masses $M_{200} (z=0) \sim 10^{12-13} M_{\odot}$. The ejected metal mass is substantial; is it similar to that remaining inside R$_{200}$ (in stars, ISM and CGM). At the same time, however, they commented that the model used for mixing the metals with pristine gas is not yet at an advanced level. This might affect %underestimate 
the amount of the metals ejected into intergalactic space in the EAGLE simulations.

In a similar vein, 
\citet{Chadayammuri:22} found that the EAGLE simulation overpredicts the X-ray emission of the stacked eROSITA %\LEt{ Consider defining.}
data of a sample of over 2000 galaxies in the central 100 kpc region for galaxies of stellar mass of $M_{*} \sim 10^{11} M_{\odot}$. They speculated that the AGN feedback model in EAGLE is not efficient enough in blowing the gas out. \citet{Chadayammuri:22} point out that $M_{*} \sim 10^{11} M_{\odot}$ galaxies in Illustris-TNG simulation better fit the eROSITA data, possibly because it is calibrated to better reproduce the baryon content within groups and clusters.
Consistently with the above, the EAGLE gas mass fraction in $\log M(M_{\odot}) = 13-14$ haloes within $R_{500}$ is higher than that 
in the Illustris-TNG and Magneticum simulations \citep[Fig. 2, left panel]{2021MNRAS.504.5131L}, suggesting a weaker EAGLE feedback for haloes within this mass range.

In SIMBA simulations \citep{Dave:2019}, the majority of the \ion{O}{VII} ions are located outside the virial radii of the haloes. The density peaks at  $\sim$ 2-3 $\times$ $R_{200}$ distance from the nearest halo, predominantly within the filaments of the Cosmic Web \citep{Bradley:2022}. 
This behaviour is somewhat different from what we obtain in the EAGLE simulations, where the \ion{O}{VII} density decreases beyond the virial radius. While the essential difference between the SIMBA and EAGLE simulations is the implementation of bipolar kinetic outflows from the AGN in the former, the runs without jets give almost identical results. Thus, it is not currently clear what is causing the difference. 

However, the comparison of the results from SIMBA and EAGLE simulations is consistent with the above suggestion that EAGLE underestimates the ejection of metals from haloes into the IGM. 
In this case the detection probabilities of the intergalactic \ion{O}{VII} presented in our work would be too small, as more absorbing systems would have densities above the limiting column density of $\log N(\ion{O}{VII}) > 15$ in the intergalactic space.
We will investigate this issue in future work by analysing cosmological simulations with significantly different recipes for the metal enrichment.

%-----------------------------------------------------------------

\section{Conclusions}

In this work we have analysed the distribution of the intergalactic oxygen atoms and $\ion{O}{VII}$ ions within 
%Bisous 
the filaments of the Cosmic Web in the EAGLE simulation. Since oxygen is formed in and expelled from galaxies, we also studied the surroundings of haloes.  
We characterised the spatial distributions of oxygen and $\ion{O}{VII}$ and studied their mass and volume filling fractions in the filaments. 
We used this information to estimate the fraction of hot WHIM that can potentially be traced by $\ion{O}{VII}$, as well as %\LEt{ or "traced by $\ion{O}{VII}$ and {VII} column densities" depending on your meaning.} and 
the feasibility of detecting intergalactic \ion{O}{VII} absorbers %column densities 
with Athena X-IFU. 
Here we summarise our main conclusions:

   \begin{enumerate}
    
    \item The projected areas covered by filaments detected with the Bisous formalism \citep{2014MNRAS.438.3465T} in the observational SDSS survey and in the EAGLE simulation mask over 90\% of the total area. In addition, both observations and the simulation %\LEt{ or "both observations and the simulation".} 
    agree to within 1\%, indicating that the basic geometric properties of the filaments, that is, number densities and sizes, are properly reproduced in the EAGLE simulation.
    
    \item Of all the intergalactic oxygen within Bisous filaments, $\approx 72\%$ lies above the hot WHIM temperature limit of $\log T(K) > 5.5$. Consequently, $\approx 33\%$ of the intergalactic filamentary oxygen has been ionised to \ion{O}{VII}.
 
    \item According to the EAGLE simulations, the density profiles of the intergalactic \ion{O}{VII} decline rapidly beyond the virial radius of haloes. The median radial extent of \ion{O}{VII} above the approximate Athena X-IFU detection limit is $\approx 700$ kpc when an optimal observation scenario is considered.

    \item Since galaxies are relatively far apart from one another, filaments are not fully filled with intergalactic \ion{O}{VII} at densities above the approximate Athena X-IFU detection limit. Indeed, the volume filling fraction is %of OVII above detection limit...
    $\sim 1\%$ of the total filament volume. This implies that most of the \ion{O}{VII} detectable with Athena X-IFU is located within small envelopes around haloes. 

   \item The highly inhomogeneous distribution of the detectable \ion{O}{VII} complicates the usage of the measurements of the intergalactic \ion{O}{VII} absorbers for tracing the missing baryons and estimating their contribution to the cosmic baryon budget. Nonetheless, a significant fraction of the hot WHIM  ($\approx 27\%$)  may be traced with intergalactic \ion{O}{VII}.

    \item The probability of encountering an intergalactic \ion{O}{VII} absorber with $\log N_{\ion{O}{VII}} ($cm$^{-2}) > 15$ in the filament sample obtained by \citet{2014MNRAS.438.3465T} from the optical spectroscopic SDSS galaxy survey is $10-20\%$ per sight line. The expected extension of the path lengths through filaments afforded by the advent of 4MOST surveys will improve the probability to a comfortable level of $\sim$ 50\%.

  \item  The probability of intercepting a detectable \ion{O}{VII} absorber is very sensitive to the column density. Thus, if in reality feedback processes are more effective at expelling metals into the IGM than those implemented in EAGLE, the estimates for the interception probability will be larger. Similarly, slight improvements in the \ion{O}{VII} absorption sensitivity with instruments beyond the X-IFU limit would significantly boost the probability of intercepting an absorbing system.

   \end{enumerate}

Based on EAGLE, the amount of missing baryons that is expected to be traceable with next-generation X-ray instruments is limited to the immediate vicinity of haloes. Since a significant fraction of baryons lie in the diffuse intergalactic regime, probing only the high-density regions around haloes would fall short of accounting for the whole of the missing baryons. Nonetheless, a significant fraction of the missing baryon mass ($\approx 7-27\%$) resides in the outskirts of haloes beyond the virial radius, $R_{200}$. Thus, Athena X-IFU will provide invaluable observations for reducing the amount of missing baryons.

\begin{acknowledgements}
      We would like to thank Joop Schaye for his invaluable comments. We also thank the referee for their helpful and educated suggestions.
    We acknowledge the support by the Estonian Research
Council grants PRG1006, and by the European Regional
Development Fund (TK133).  We acknowledge the Virgo Consortium for making their simulation data available. The eagle simulations were performed using the DiRAC-2 facility at Durham, managed by the ICC, and the PRACE facility Curie based in France
at TGCC, CEA, Bruyèresle-Châtel.
\end{acknowledgements}

\bibliographystyle{aa}
\bibliography{eaglebib}

\end{document}